\documentclass[pdflatex,sn-mathphys-num,iicol,remarkboxoff]{sn-jnl}
\usepackage[T1]{fontenc}
\usepackage[utf8]{inputenc}
\usepackage{amsmath,amssymb,booktabs,graphicx,xcolor,tabularx}
\usepackage[title]{appendix}
\usepackage{microtype}
\usepackage{placeins}
\usepackage[noautobase,nospread]{flushend}

\hypersetup{pdftitle={Continuous Universal Fake Factors for jet-to-tau background estimation},pdfauthor={Lei Guo, Antonio De Maria, Chengxin Liao, Zihang Liu, Wenyi Jin, Jiarong Yuan, Yuchen Cai, Da Xu, Xuai Zhuang, Yanjun Tu, Yang Liu},pdfsubject={}}
\newcommand{\Azero}{A_0}
\newcommand{\FF}{\mathrm{FF}}
\newcommand{\cuff}{\mathrm{CUFF}}
\begin{document}
\title[Continuous Universal Fake Factors]{Continuous Universal Fake Factors for jet-to-tau background estimation}
\author[1]{\fnm{Lei}\sur{Guo}}
\author[2]{\fnm{Antonio}\sur{De Maria}}
\author[3,4]{\fnm{Chengxin}\sur{Liao}}
\author[1]{\fnm{Zihang}\sur{Liu}}
\author[3,4]{\fnm{Wenyi}\sur{Jin}}
\author[3,4]{\fnm{Jiarong}\sur{Yuan}}
\author[5,6]{\fnm{Yuchen}\sur{Cai}}
\author[3,4]{\fnm{Da}\sur{Xu}}
\author[3,4]{\fnm{Xuai}\sur{Zhuang}}
\author[7]{\fnm{Yanjun}\sur{Tu}}
\author*[1]{\fnm{Yang}\sur{Liu}}
\email{liuy2399@mail.sysu.edu.cn}
\affil[1]{School of Science, Shenzhen Campus of Sun Yat-sen University, Shenzhen, Guangdong, 518107, China}
\affil[2]{Nanjing University, Nanjing, China}
\affil[3]{Institute of High Energy Physics, Chinese Academy of Sciences, Beijing 100049, China}
\affil[4]{University of Chinese Academy of Sciences, Beijing 100049, China}
\affil[5]{INFN Sezione di Bologna, Bologna, Italy}
\affil[6]{Dipartimento di Fisica e Astronomia A. Righi, Università di Bologna, Bologna, Italy}
\affil[7]{Department of Physics, The University of Hong Kong, Hong Kong SAR, China}
\abstract{We study Continuous Universal Fake Factors (CUFF), a method for estimating backgrounds from jets misidentified as hadronic tau candidates.
CUFF combines four reference-region fake factors for each target tau candidate failing Loose identification.
Their weights depend on candidate features and target fail-Loose distributions after subtracting simulated contributions from genuine taus and electrons or muons misidentified as tau candidates.
We use ATLAS 2015--2016 Open Data with tau candidates matched to jets with $20<p_T^{\rm jet}\leq60$ GeV.
In eight mixtures of reference samples with known proportions, predicted total yields for the Medium identification selection differ from data after subtraction by $+0.7\%$ to $+4.3\%$, supporting the combination.
In two validation regions, the corresponding differences are $-21.85\%$ and $-45.84\%$, with statistical uncertainties, including training fluctuations, of $1.58\%$ and $6.03\%$ relative to the data references.
These discrepancies lie within broad empirical uncertainty ranges from simulation comparisons between regions, widened by limited simulation statistics.
For Loose identification, CUFF and the Universal Fake Factor (UFF) method adapted to public data give similar inclusive yields in one validation region, while CUFF agrees more closely with the data reference in the other.
CUFF also describes the shapes of several kinematic distributions more accurately than adapted UFF, including that of an observable not used as a model input.
Across nine identification thresholds, relative yield differences vary gradually from $-26.4\%$ to $-19.7\%$ and $-50.2\%$ to $-40.6\%$ in the two regions, respectively, demonstrating consistent application of CUFF at continuously variable tau-identification thresholds without retraining.}
\onecolumn
\maketitle
\twocolumn

\section{Introduction}
\label{sec:introduction}
The tau lepton is the heaviest charged lepton in the Standard Model. Its short
lifetime means that it is observed through its decay products. It decays
to an electron or muon and two neutrinos in approximately 35\% of cases,
and to hadrons and a tau neutrino in the remaining approximately 65\%.
Hadronic decays usually contain one or three charged particles, often
accompanied by neutral pions~\cite{PDGTau,HiggsTau}. The visible products
of these hadronic tau decays form a narrow detector signature, while the neutrinos carry away
unmeasured momentum.

Tau-lepton final states provide important probes of Standard Model (SM)
interactions and physics beyond the SM (BSM). Measurements of Higgs-boson
production in the $H\to\tau^+\tau^-$ decay channel probe the Higgs
Yukawa coupling to tau leptons~\cite{HiggsTau}. Searches for charged Higgs bosons
through $H^+\to\tau^+\nu_\tau$, with a tau lepton and a tau neutrino in
the final state, test extensions of the SM Higgs sector~\cite{ChargedHiggsTau}.

In analyses with hadronic taus, quark- or gluon-initiated jets
misidentified as hadronic tau candidates constitute an important
background that must be estimated accurately. Predicting this jet-to-tau
background with Monte Carlo (MC) simulation is difficult because the
misidentification rate depends on jet fragmentation and the detector
response. Small misidentification rates also require large simulated
samples. Data-driven methods are therefore widely used to estimate this
contribution~\cite{ChargedHiggsTau,UFF}.

The fake-factor (FF) method is among the most widely used data-driven
approaches to estimating fake-tau backgrounds~\cite{FakeLeptonTools,UFF,NFF}.
It predicts the jet-to-tau background yield in the target ID region by
weighting candidates in the corresponding anti-ID region with pass-to-fail
ratios measured in dedicated control regions.

\begin{figure}[t]
\centering
\includegraphics[width=\linewidth]{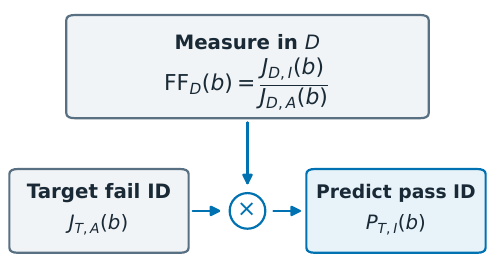}
\caption{Traditional FF measurement and application in one kinematic bin
$b$. The yields $J$ have non-jet contamination subtracted. The ratio of
passing ($I$) to failing ($A$) yields in the dedicated control region $D$ weights the failing target
sample in $T$ to predict its passing jet-fake yield}
\label{fig:traditional-ff}
\end{figure}

Figure~\ref{fig:traditional-ff} illustrates the FF measurement and
application. Here, $D$ denotes the dedicated control region, $T$ the target
region, and $b$ a kinematic bin. The labels $I$ and $A$ indicate
candidates passing and failing the chosen tau-identification requirement,
respectively. The yields
$J$ are obtained from data after subtracting MC estimates of genuine taus
and electrons or muons misidentified as tau candidates; these contributions
are termed non-jet contamination. The measured FF,
$\FF_D(b)$, is applied to $J_{T,A}(b)$ to obtain the predicted jet-to-tau
yield $P_{T,I}(b)$ in the target ID region.

The main sources of jet-to-tau backgrounds considered here are light-quark
jets, $b$-quark jets, gluon jets and jets from pileup interactions. These
sources have different probabilities of passing tau identification, so
the FF depends on their relative contributions in the anti-ID region.
Different target regions can therefore require different FFs even at the
same tau $p_T$ and track multiplicity. Conventional FF applications often
use dedicated control regions selected to reproduce the target
composition. Defining such regions with sufficient jet-fake purity and
statistical precision can be challenging. Limited statistics also
restrict the number of variables and bins used to parameterize the
FF~\cite{UFF,NFF,FakeLeptonTools}.

\begin{figure}[t]
\centering
\includegraphics[width=\linewidth]{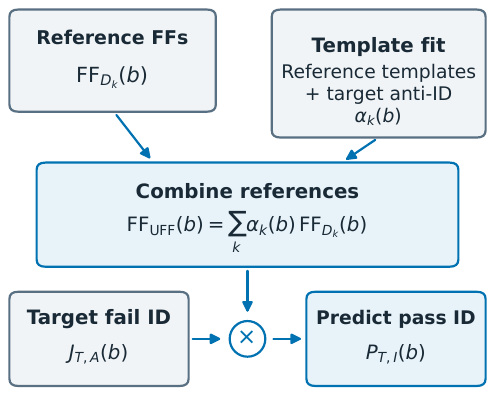}
\caption{UFF combination in one kinematic bin $b$. Four reference
measurements supply the FFs and templates. Fitting those templates to
the target anti-ID sample determines $\alpha_k(b)$, which combine the
reference FFs. Non-jet contributions are accounted for in the measurements
and the target application}
\label{fig:uff-overview}
\end{figure}

To enable the use of common FF measurements across different analyses,
the ATLAS Collaboration developed the Universal Fake Factor (UFF) method~\cite{UFF}. Its
four dedicated control regions $D_k$ ($k=1,\ldots,4$), referred to below as
reference regions, enrich light-quark, $b$-quark, gluon and pileup
jet fakes, while each remains a mixture of sources. The reference
selections comprise control regions enriched in $Z\to\mu^+\mu^-$ and
$t\bar t$ events, together with two multijet control regions separated
by a jet-vertex-tagging requirement~\cite{JVT}. In each tau-$p_T$
and track-multiplicity bin, UFF forms
\begin{equation}
 \FF_{\rm UFF}(b)=\sum_{k=1}^{4}\alpha_k(b)\,\FF_{D_k}(b).
 \label{eq:intro-uff}
\end{equation}
The coefficients $\alpha_k$ are obtained by fitting reference templates
of a quark/gluon discriminant~\cite{QGTagger} to the target anti-ID distribution. The
fitted coefficients weight the four reference FFs in each kinematic bin
and can be negative. The reference FFs and templates are shared across
analyses, while the combination coefficients are determined separately
for each target region by fitting its anti-ID distribution.
Figure~\ref{fig:uff-overview} summarizes this measurement and combination.

In the published UFF implementation, reference FFs and combination
coefficients are determined in bins of tau $p_T$ and track multiplicity.
Candidates in the same bin of a given target region therefore receive
the same combined FF, even when other properties relevant to tau
identification differ. Finer binning or additional binning variables
divide the available data from the dedicated control regions into more
bins, limiting statistical precision when counts become small~\cite{UFF,NFF}.
Changing the Tau-ID
working point also requires corresponding FF and template inputs.
These limitations motivate a multivariable formulation with
candidate-dependent FFs and reference weights, in which the FFs depend
continuously on the identification threshold. The objective is to evaluate
the same trained models at different thresholds within the supported range.

We propose the Continuous Universal Fake Factors
(CUFF) method, combining the reference-region approach of UFF with multivariable
learning and continuous Tau-ID evaluation. Neural networks learn the FFs
in the reference regions from data candidates passing and failing Loose
identification, corrected for non-jet contamination. For reference region
$k$, let $x$ denote the FF model inputs and $t_L$ the candidate's Loose
threshold. The FF at a threshold $t$, with $t_L\leq t\leq1$, is
\begin{equation}
 \FF_k(x;t)=\FF_k(x;t_L)\,S_k(t\mid x).
 \label{eq:intro-cuff-factorization}
\end{equation}
Here $\FF_k(x;t_L)$ is the Loose-to-fail-Loose fake factor at
fixed $x$, and $S_k$ is the fraction of Loose-ID jet-fake candidates that
also pass $t$:
\begin{equation}
 S_k(t\mid x)=\int_t^1 p_k(s\mid x,s>t_L)\,\mathrm{d}s.
 \label{eq:intro-cuff-tail}
\end{equation}
The variable $s\in[0,1]$ is the Tau-ID score, and $p_k$ is its normalized
probability density for Loose-ID jet-fake candidates at fixed $x$.
The model learns the Loose FF and this score distribution; changing $t$
then requires only evaluating the integral at the new threshold, without
retraining.

The common application sample $\Azero$ consists of tau candidates in the
target region that fail Loose identification. Its non-jet contamination
is estimated using MC simulation. For signal selections requiring Loose
or tighter Tau-ID, these candidates lie outside the ID selection. Data
in the target ID region are not used to construct the prediction. Keeping
$\Azero$ fixed also allows the same application sample and FF denominator
to be used as the identification threshold changes. Compared with a
working-point-specific anti-ID sample, this choice does not use target
candidates that pass Loose but fail a tighter working point, so it does
not minimize the statistical uncertainty at each threshold.
For analyses requiring customized Tau-ID working points, this flexibility
enables continuous threshold optimization beyond the standard choices,
without new FF measurements or retraining within the modelled score range.

A separate neural network combines the reference FFs with learned
weights $g_k$~\cite{MoE}. It is trained using reference-region labels and
mixtures with known proportions. When applied to a target, the weights
depend on several candidate and event observables and on a summary of the
target $\Azero$ data after non-jet subtraction. The summary is independent
of the order of events~\cite{DeepSets}. The weights can vary between
candidates in the same target region and are not restricted to a single
q/g discriminant. No new target template fit is required, and the same
weights are retained across Tau-ID thresholds.

This multivariable description can resolve dependencies averaged over by
low-dimensional binning. CUFF may also benefit differential measurements
that use unfolding.

Previous studies have demonstrated the feasibility of learning FFs from
data. Neural Fake Factors (NFF)~\cite{NFF} use classifiers to estimate distribution
ratios after correcting for genuine-lepton contamination; their ATLAS
Open Data example predicts fake-electron backgrounds in a $W\to e\nu$
selection~\cite{ClassifierRatio}. CMS MUFFIN~\cite{MUFFIN} provides a hadronic-tau
example using boosted decision tree reweighting and process-dependent
transfer factors. Building on these developments, this study
focuses on learning the reference combination and providing predictions
at continuously variable Tau-ID thresholds within one procedure.

\begin{figure*}[tp]
\centering
\includegraphics[width=\linewidth]{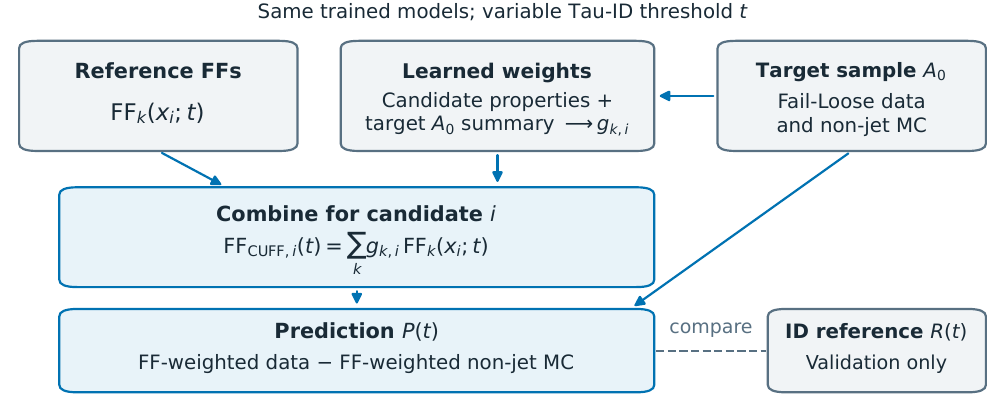}
\caption{Application of the trained CUFF models. The reference FFs,
$\FF_k(x_i;t)$, and weights, $g_k(v_i,C_T)$, are evaluated for each target
fail-Loose data or non-jet-MC candidate. Their weighted sums give the
prediction $P(t)$ after non-jet subtraction. The ID data determine the
comparison yield $R(t)$. The target sample and combination weights are
retained when the requested threshold changes}
\label{fig:method-overview}
\end{figure*}

We test the individual FFs, their combination in known reference mixtures,
and their application to two validation regions using ATLAS
2015--2016 Open Data. Nine identification thresholds examine reuse of the trained models,
while a comparison with UFF adapted to public inputs tests yields and
distribution shapes, including observables not supplied to the models.
These tests establish the scope of this proof of concept. The CUFF
procedure is illustrated in Figure~\ref{fig:method-overview}.

The rest of this paper is organized as follows.
Section~\ref{sec:data} describes the data and event selections.
Section~\ref{sec:method} presents the CUFF implementation and component
checks. Section~\ref{sec:results} presents the validation tests and
results. Section~\ref{sec:uncertainties} describes the uncertainty
estimation, and Section~\ref{sec:discussion} summarizes the study and its
outlook. The appendices contain technical definitions and supplementary
validation details.

\section{Data samples and event selection}
\label{sec:data}
This study uses ATLAS Open Data for Research from 2015--2016 in
DAOD\_PHYSLITE format~\cite{OD2015,OD2016}, corresponding to an integrated
luminosity of approximately $36.6\,\mathrm{fb}^{-1}$.

Truth information in MC identifies the non-jet contributions: genuine
taus decaying leptonically and hadronically, and electrons or muons misidentified as tau
candidates. Their predicted yields are subtracted from data in both the
ID and fail-Loose regions. The subtraction acts on yields and
distributions; it does not assign truth labels to individual data
candidates. It also accounts for genuine taus that fail Loose
identification. Simulated candidates classified as jet fakes are used in separate
transfer tests and are excluded from the non-jet subtraction. The nominal
FF training remains based on data. Simulated candidates without an
assigned truth category are retained as unclassified.

MC event weights include the signed generator weight, the stored pileup
weight and normalization to the integrated luminosity of the corresponding
trigger selection. Additional efficiency scale factors
are not applied in this Open Data study, and their associated
uncertainties are not estimated.

\subsection{Candidate and event definitions}
The object selections described below are applied to both the reference
and validation regions.
Varying the tau-identification threshold leaves this baseline selection fixed,
so all thresholds use the same reconstructed candidate sample. Events satisfy
the public good-run list and jet-cleaning requirements, and have
a unique stored primary vertex with at least two track links.
Object four-momenta are taken from the public data.
Object overlap removal and the missing transverse momentum calculation use
public ATLAS analysis tools. The objects entering the
missing-momentum calculation remain fixed across identification thresholds.

Muons have $p_T>7$ GeV and $|\eta|<2.5$, satisfy Loose identification~\cite{MuonRecoRun2}
and pass the stored inner-detector track-quality requirements. Electrons
have $p_T>15$ GeV, satisfy Loose likelihood identification~\cite{ElectronRecoRun2} and pass the
calorimeter-cluster quality requirements. Their cluster pseudorapidity
at the second electromagnetic calorimeter layer, $\eta_{\rm cl}$, satisfies
$|\eta_{\rm cl}|<1.37$ or $1.52<|\eta_{\rm cl}|<2.47$.
Muon and electron tracks have transverse impact-parameter significance
$|d_0|/\sigma(d_0)<3$ and $<5$, respectively, and
$|z_0^{\rm PV}\sin\theta|<0.5$ mm. Here $d_0$ is the transverse impact
parameter, $\sigma(d_0)$ includes the beam-spot uncertainty,
$z_0^{\rm PV}$ is the longitudinal impact parameter relative to the
selected primary vertex, and $\theta$ is the track polar angle.
Tag muons satisfy a variable-cone track-isolation ratio below $0.15$ and
a calorimeter-isolation ratio below $0.30$; the tag-electron ratios are $0.15$
and $0.20$~\cite{MuonRecoRun2,ElectronRecoRun2}.

Exactly one tau candidate is required after overlap removal. It has
$p_T>20$ GeV, $|\eta|<1.37$ or $1.52<|\eta|<2.5$, unit absolute charge, and
one or three charged tracks (one- or three-prong). The transformed
DeepSet score~\cite{TauDeepSetConfig} must be finite and in $[0,1]$.
No minimum Tau-ID score is required at this stage.
One-prong candidates satisfy the stored Tight electron veto based on a
recurrent neural network (RNN)~\cite{TauRun3Performance}. Tau candidates
within $\Delta R<0.2$ of retained electrons, muons or photons are removed.

Each tau candidate is associated with the nearest particle-flow jet before
tau--jet overlap removal, within
$\Delta R<0.4$, with $p_T>20$ GeV and $|\eta|<2.5$.
The FF is applied to the tau candidate, with additional input observables
from its associated jet. A valid association and
$p_T^{\mathrm{associated\ jet}}\leq60$ GeV are required in every reference
and validation region. This common momentum range is chosen for the
NNJvt-based separation of the multijet references, motivated by the default
range of the public NNJvt selection tool~\cite{NNJvtTool}.
The use of the stored \texttt{NNJvtPass} decision is detailed in
Appendix~\ref{app:nnjvt-input}.

Jets used to determine the jet and $b$-jet multiplicities are selected
from the same particle-flow jet collection~\cite{ParticleFlowJets}. They have $p_T>20$ GeV and
$|y|<4.5$, pass the applicable central
NNJvt and forward-JVT requirements and electron, muon and photon overlap removal,
and are separated from the tau candidate by at least $0.2$ in $\Delta R$.
The $b$-jet count uses these jets with $|\eta|<2.5$ and the public
DL1dv01 implementation of the DL1d tagger~\cite{DL1dTagging} at the 85\% operating point.

The selections use single-muon, single-electron and jet triggers,
following the tag-object choices in UFF. For each year and object category,
the triggers are combined with a logical OR. Only active, unprescaled triggers are
used to select data events. The selected tag lepton must match an object associated
with a fired lepton trigger.

\subsection{Reference and validation regions}
The reference FFs are measured in a $W$ region with same-sign
(SS) muon--tau pairs, together with the $t\bar t$, MJ-pass and MJ-fail reference regions. These regions provide
complementary mixtures of jet-fake sources. The $W$ selection targets
quark enrichment, while the electron--muon $t\bar t$ selection targets
heavy-flavour enrichment. The multijet selections use the associated
jet's NNJvt decision to target gluon-enriched and pile-up-enriched samples,
following the roles of the high- and low-JVT references in UFF~\cite{UFF}.
Each reference region contains a mixture of fake sources.

Compared with the published UFF implementation, we replace the
$Z(\mu\mu)$ reference with a same-sign $W(\mu\nu)$ region to provide a
quark-enriched reference for the W-like validation regions.
This choice is motivated by the regional differences in quark-fake factors
discussed in UFF and its recommendation to consider W-based inputs for
quark-fake-dominated analyses sensitive to those differences~\cite{UFF}.

Two validation regions are defined to test the CUFF background prediction
in different jet environments. Both are based on the $W(\mu\nu)$ selection
but require opposite-sign (OS) muon--tau pairs, separating them from the
same-sign W reference region. The $N_b=0$ region requires no selected
$b$ jets, whereas the $\mu3j$ region requires at least three selected jets
and at least one $b$ jet. These requirements make the two validation
regions mutually exclusive.
The reference and validation selections are summarized in
Table~\ref{tab:regions}.

\begin{table*}[t]
\centering\small
\caption{Reference regions used to determine the FF models and validation
regions used to test the CUFF prediction. All rows satisfy the common tau-candidate and
associated-jet requirements. Tau-ID pass/fail selections are applied in
addition to these requirements. The two validation regions are mutually
exclusive.}
\label{tab:regions}
\begin{tabularx}{\linewidth}{lX}
\toprule
Sample & Additional selection requirements \\
\midrule
SS $W$ reference & Exactly one retained post-overlap baseline muon, required
to be isolated and Medium-or-better with $p_T>28$ GeV,
same-sign tau, $\Delta R(\mu,\tau)>0.4$, electron/photon veto,
$E_T^{\rm miss}>30$ GeV and $40<m_T(\mu,E_T^{\rm miss})<100$ GeV;
single-muon trigger. No additional $b$-jet or jet-multiplicity requirement. \\
$t\bar t$ reference & One isolated electron satisfying Tight identification with $p_T>28$ GeV and
one isolated Medium-or-better muon with $p_T>10$ GeV, opposite sign;
at least one $b$ jet, photon veto; single-electron trigger. \\
MJ-pass and MJ-fail references & Electron, muon, photon and $b$-jet vetoes; at least two
selected jets with leading $p_T>390$ (450) GeV in 2015 (2016),
$|\eta_{j1,j2}|<3$, $\Delta\phi_{j1,j2}>2.7$, and
$p_T^\tau<(p_T^{j1}+p_T^{j2})/8$. No third-jet veto in the accepted relaxed
selection. The associated-jet NNJvt decision separates pass and fail. \\
OS $N_b=0$ validation & The SS $W$ object, trigger, $E_T^{\rm miss}$ and $m_T$
requirements, with opposite muon--tau charge and no selected $b$ jet. \\
OS $\mu3j$ validation & The same opposite-sign selection, with at least three
selected jets and at least one selected $b$ jet. \\
\bottomrule
\end{tabularx}
\end{table*}

\section{Implementation of CUFF}
\label{sec:method}
CUFF is implemented with four networks that determine the reference FFs
and two networks that determine their combination weights. The reference
data provide the training samples and the component checks. The trained
models are then applied to the target fail-Loose sample at the requested
Tau-ID threshold.

\subsection{Determination of the reference fake factors}
\label{sec:reference-ff}
\label{sec:reference-training}
\label{sec:reference-checks}
A neural network is trained in each of the four reference regions to
estimate its FF from data candidates passing and failing Loose
identification. The training proceeds in two stages: correction of the
training data for non-jet contamination, followed by learning the
Loose-to-fail-Loose ratio and the score distribution of passing candidates.
Table~\ref{tab:features} lists the inputs of the correction and FF networks.

\begin{table*}[tp]
\centering\small
\caption{Input variables of the CUFF models. Model-specific additions are
listed below the common input set.}
\label{tab:features}
\renewcommand{\arraystretch}{1.08}
\begin{tabularx}{\linewidth}{lcX}
\toprule
\multicolumn{3}{l}{\textit{Common inputs shared by the models listed below}} \\
\midrule
Group & Number & Variables \\
\midrule
Tau candidate & 3 & $p_T^\tau$, $|\eta^\tau|$, and the one-/three-prong category. \\
Associated jet & 4 & $p_T$, track multiplicity, track width, and the track-based shape variable $C_1$. \\
Flavour tagging & 2 & Associated-jet probability ratios $p_b/p_u$ and $p_c/p_u$. \\
Pile-up & 1 & Average number of interactions per bunch crossing, $\langle\mu\rangle$. \\
Missing-value indicators & 4 & Flags for unavailable or invalid associated-jet track multiplicity, track width, $C_1$, and flavour-tagger probabilities. \\
\midrule
Common input total & 14 & \\
\addlinespace[0.7em]
\multicolumn{3}{l}{\textit{Complete inputs for each model}} \\
\midrule
Model & Total & Additions to the common inputs \\
\midrule
Non-jet-correction network (full) & 17 & $E_T^{\rm miss}$, $N_{\rm jets}$, and the Tau-ID score through $\log[s/(1-s)]$. \\
MJ-fail FF & 14 & None. \\
MJ-pass FF & 16 & $E_T^{\rm miss}$ and $N_{\rm jets}$. \\
$t\bar t$ FF & 17 & $E_T^{\rm miss}$, $N_{\rm jets}$, and $p_T^{j_3}$. \\
SS $W$ FF & 19 & $E_T^{\rm miss}$, $N_{\rm jets}$, $p_T^{j_3}$, the associated-jet NNJvt decision, and $p_T^{j_1}$. \\
Reference classifier & 16 & $E_T^{\rm miss}$ and $N_{\rm jets}$. \\
Reference-weight network & 37 & $E_T^{\rm miss}$, $N_{\rm jets}$, $p_T^{j_3}$, the associated-jet NNJvt decision, and $p_T^{j_1}$; four classifier log-probabilities, $\log q_k^{(0)}$; and the 14 entries of the target summary $C_T$. \\
\bottomrule
\end{tabularx}
\par\smallskip
\begin{minipage}{\linewidth}
\footnotesize
The prong category is encoded by a binary flag. The flavour-tagger outputs
$p_b$, $p_c$ and $p_u$ correspond to $b$, $c$ and light-flavour jets,
respectively. $N_{\rm jets}$ counts the selected jets defined in
Section~\ref{sec:data}; $j_1$ and $j_3$ are their leading and third-leading
jets. The reference-weight network has 19 candidate/event inputs, four
classifier log-probabilities and 14 target-summary values.
Logarithmic transformations and scalings follow the model configurations.
The reduced contamination network uses only the prong category and
score logit; its selection and the constant corrections are described in
Appendix~\ref{app:contamination-implementation}.
\end{minipage}
\end{table*}

The non-jet contribution is estimated using MC. Each data candidate's
contribution to the training loss is multiplied by a correction weight,
so candidates with a larger estimated non-jet fraction contribute less.
The correction is determined separately in $\Azero$ and Loose-ID samples.
For sufficiently large samples, a classifier is trained to distinguish data from
non-jet MC, with event weights applied during training. The 17 input variables, denoted by
$z_c$, comprise the
14 common variables in Table~\ref{tab:features}, $E_T^{\rm miss}$,
$N_{\rm jets}$ and the Tau-ID score through $\log[s/(1-s)]$.
With the data and non-jet-MC yields per unit $z_c$ denoted by $\rho^{\rm Data}_k$
and $\rho^{\rm non\text{-}jet,MC}_k$, the estimated correction weight is
\begin{equation}
 c_k(z_c)=1-\frac{\rho^{\rm non\text{-}jet,MC}_k(z_c)}
                       {\rho^{\rm Data}_k(z_c)}.
 \label{eq:contamination}
\end{equation}
The network output is restricted to give weights between zero and one.
Samples with insufficient data or MC statistics use a two-input network
based on the prong category and the same score input, or a constant
correction from the total data and MC yields.
For one-prong Loose-ID candidates in MJ-fail, separate constant weights
are determined for candidates with and without a valid associated-jet
width, because their contamination fractions differ and the non-jet-MC
statistics limit a more detailed correction. The criteria and numerical treatment are given in
Appendix~\ref{app:contamination-implementation}.

Events in the reference regions are divided into five subsets, or folds.
For each evaluation fold $E$, three folds form the training sample $T$
and the remaining fold forms the model-selection sample $V$.
The same assignments are used at every Tau-ID threshold.
Within $T$, each fold's correction weights are estimated from the other
two folds. Within those folds, one part trains the correction network
and another selects the saved parameters with the lowest validation loss. The correction weight for each candidate in $V$ is the average of its
three estimates. This gives each FF-training candidate a correction determined
from other events. The precise assignments are retained in
Appendix~\ref{app:training}.

\begin{figure*}[tp]
\centering
\includegraphics[width=\linewidth]{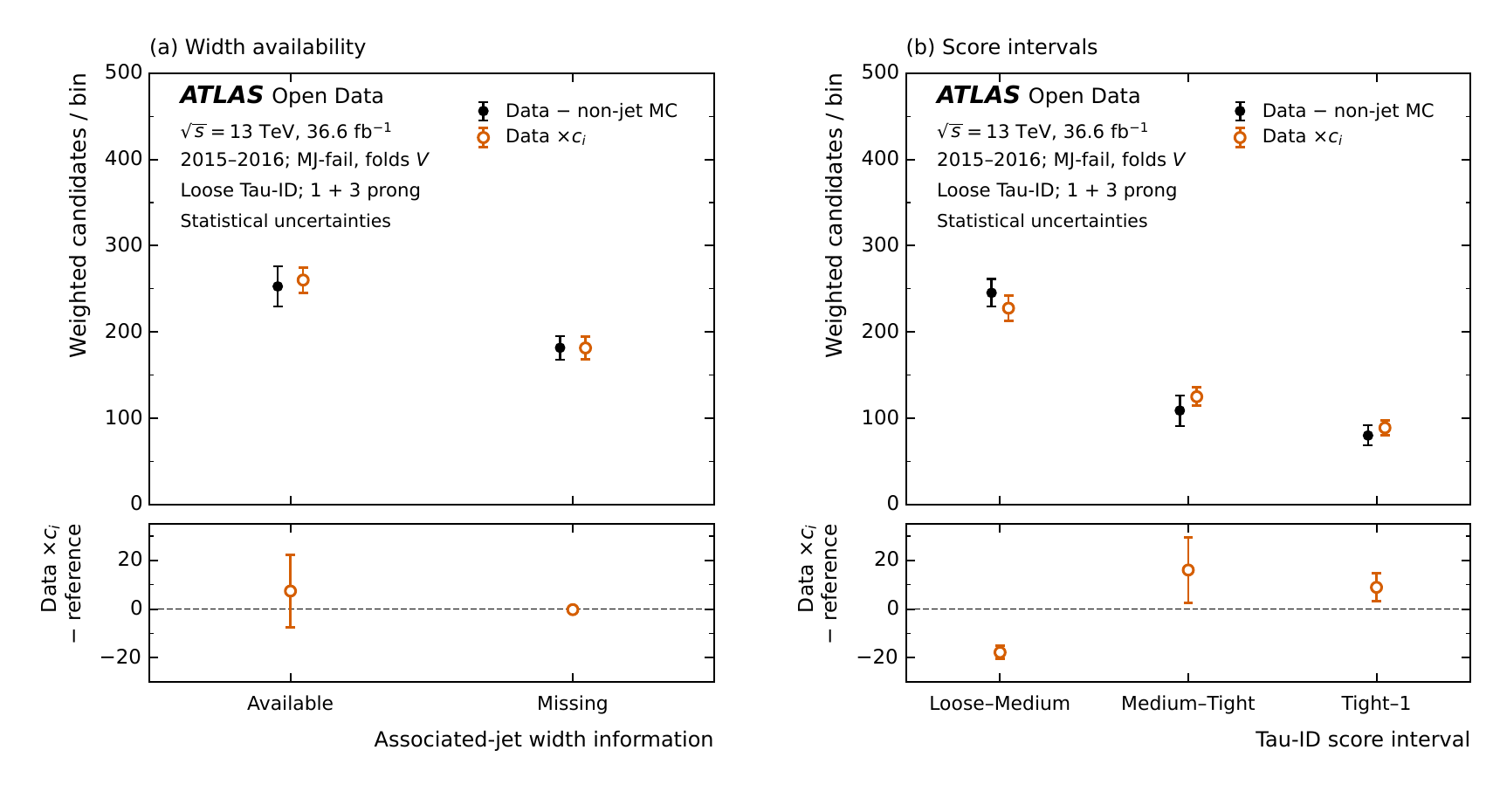}
\caption{Check of the non-jet correction used for MJ-fail training,
combining both prong categories in folds $V$. Data weighted by $c_i$ are
compared with the reference yield $R$ from data after non-jet-MC subtraction.
The left panel separates
candidates passing Loose identification according to whether a valid
associated-jet width is available; the right panel uses mutually exclusive
Tau-ID score intervals. The lower panels show $C-R$, accounting
for the shared data. Error bars show statistical uncertainties from the
finite data and MC samples; uncertainty in determining the correction
weights and in MC modelling is not included}
\label{fig:nonjet-correction}
\end{figure*}

For a specified selection, either $\Azero$ or ID, the reference yield is
$R=N^{\rm Data}-N^{\rm non\text{-}jet,MC}$.
The non-jet correction is checked by comparing the weighted data yield
$C=\sum_{i\in\mathrm{Data}}c_i$ with $R$ in the same selection.
Table~\ref{tab:nonjet-W} shows the inclusive SS $W$ comparison.
The differences are consistent with zero
within the displayed statistical uncertainties.
This agreement supports using the corrected SS $W$ data for subsequent
FF training.
Figure~\ref{fig:nonjet-correction} shows the corresponding MJ-fail
comparison. Among candidates passing Loose identification, the weighted
and subtracted yields agree within the displayed statistical uncertainties,
separately for candidates with and without a
valid associated-jet width. The
Loose--Medium score interval retains a difference of
$-17.80\pm2.78$ candidates, approximately $-7.3\%$ of $R$.
The statistical uncertainties account for the data events shared by the
two estimates. Uncertainty in determining the correction weights and in
MC modelling is not included in this check.

\begin{table}[!htbp]
\centering\small
\caption{Inclusive non-jet-correction check in the SS $W$ reference region,
combining both prong categories in folds $V$.
Central values are in percent and uncertainties in percentage points.
Statistical uncertainties account for the shared data; uncertainty in
determining the correction weights and in MC modelling is not included.}
\label{tab:nonjet-W}
\begin{tabular*}{\columnwidth}{@{\extracolsep{\fill}}lc@{}}
\toprule
Selection & $100(C-R)/R$ [\%] \\
\midrule
$\Azero$ & $ +0.021 \pm 0.036 $ \\
Loose & $ +0.095 \pm 0.169 $ \\
Medium & $ +0.181 \pm 0.235 $ \\
Tight & $ +0.110 \pm 0.355 $ \\
\bottomrule
\end{tabular*}
\end{table}

The networks used to determine the FFs have two fully connected hidden
layers with SiLU activation functions. Each layer contains 64 nodes for
SS $W$ and MJ-pass, and 16 for $t\bar t$ and MJ-fail.
For candidate inputs $x$, the outputs specify the positive Loose FF
$L_k(x)\equiv\FF_k(x;t_L)$ and eight nonnegative coefficients summing to
one. The coefficients combine eight fixed Beta distributions to describe
the passing-score distribution after rescaling the interval above Loose
to the unit interval. The observed score is a training
response, while $x$ contains the other variables listed in
Table~\ref{tab:features}.

The network is trained to determine both outputs by minimizing a weighted
negative log-likelihood.
The pass/fail term uses all candidates and constrains $L_k$ through the
passing probability $L_k/(1+L_k)$. The score-density term uses the
Loose-ID candidates. Each contribution is multiplied by $c_k(z_c)$.
AdamW~\cite{AdamW} updates the parameters using $T$; the parameters with the lowest
loss on $V$ are retained. The FF at a requested threshold follows by
integrating the learned score distribution as in
Eqs.~\eqref{eq:intro-cuff-factorization} and~\eqref{eq:intro-cuff-tail}.
The likelihood, density parametrization and optimizer settings are given
in Appendix~\ref{app:training}.

To improve the MJ-fail Loose-yield prediction, the learned FF is
calibrated in $p_T$/prong bins of the reference training sample $T$.
The corrected Loose yield is divided by the prediction from
the corrected $\Azero$ candidates, and the ratio multiplies the FF at
all thresholds in that bin. Bins that do not meet the calibration
requirements retain the original FF.
The score distribution at fixed inputs is preserved.
Further calibration details are given in
Appendix~\ref{app:mjfail-calibration}.

\begin{figure*}[tp]
\centering
\includegraphics[width=\linewidth]{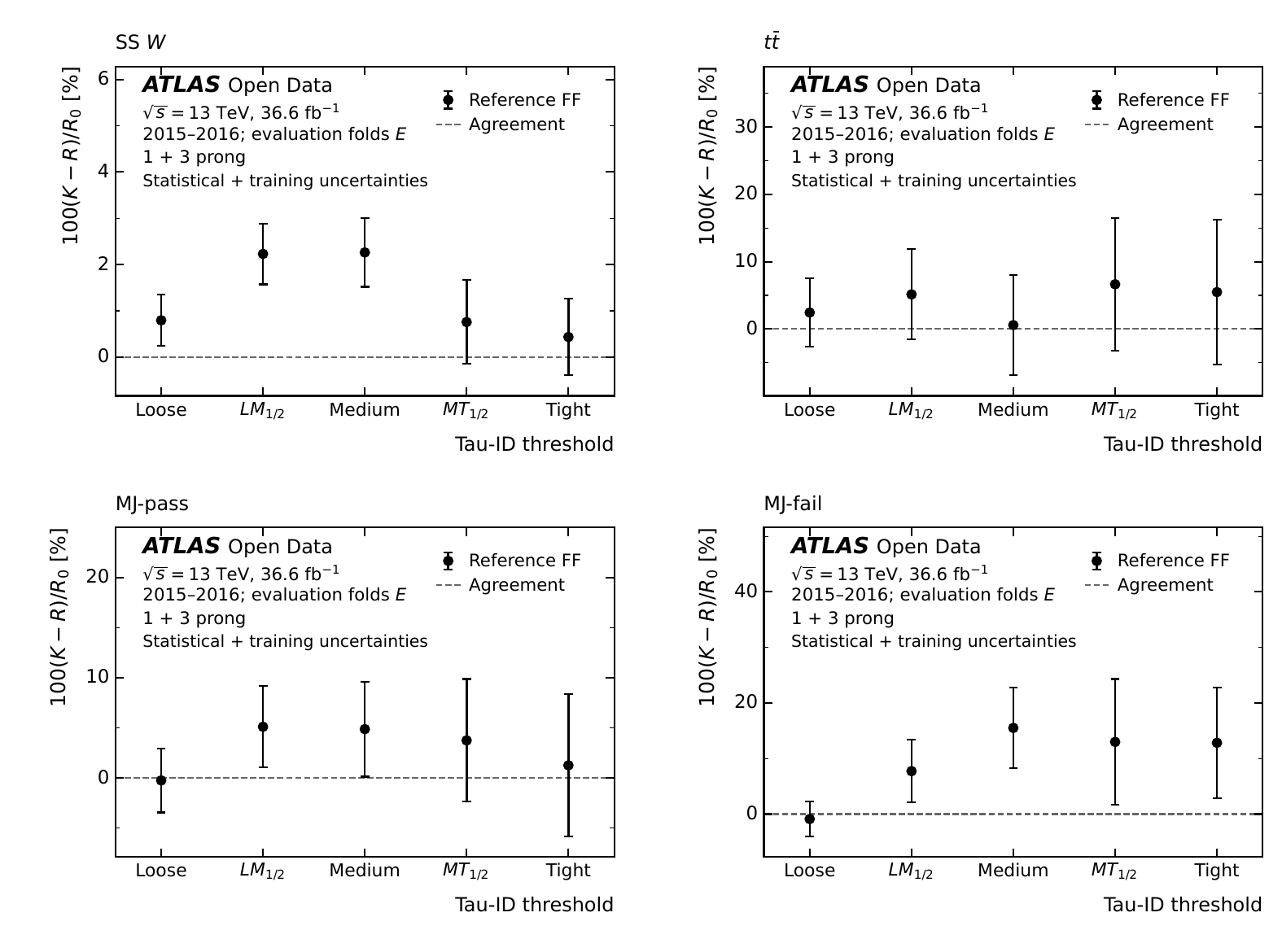}
\caption{Reference-region FF checks at the five Tau-ID
thresholds, combining one-prong and three-prong candidates. Each panel
shows $100(K-R)/R_0$ for one reference region, where $K$ is its FF-weighted
$\Azero$ data prediction after weighted non-jet-MC subtraction and $R$ is
the ID data yield after non-jet-MC subtraction. The denominator $R_0$ is
fixed to its nominal value. Error bars show statistical and training
uncertainties (one standard deviation); the thresholds share events and
trained models}
\label{fig:reference-closure}
\end{figure*}

The FFs are tested on the respective evaluation folds $E$.
For each reference, the predicted ID yield $K$ is the sum of its FFs over
$\Azero$ data, minus the correspondingly FF-weighted non-jet-MC yield.
The comparison yield $R$ is the ID data yield after non-jet-MC subtraction.
Figure~\ref{fig:reference-closure} shows the relative difference
$d=100(K_0-R_0)/R_0$ at five Tau-ID thresholds: Loose, Medium and Tight,
and the midpoints of the Loose--Medium and Medium--Tight score intervals
in each prong category. The subscript 0 denotes the nominal result;
Table~\ref{tab:source} gives the results for Medium identification.
Error bars include statistical and training uncertainties. They are
obtained from the variation of $K-R$ when the contamination correction
and FF training are repeated using statistically resampled data and MC
samples.

\begin{figure*}[tp]
\centering
\includegraphics[width=\linewidth]{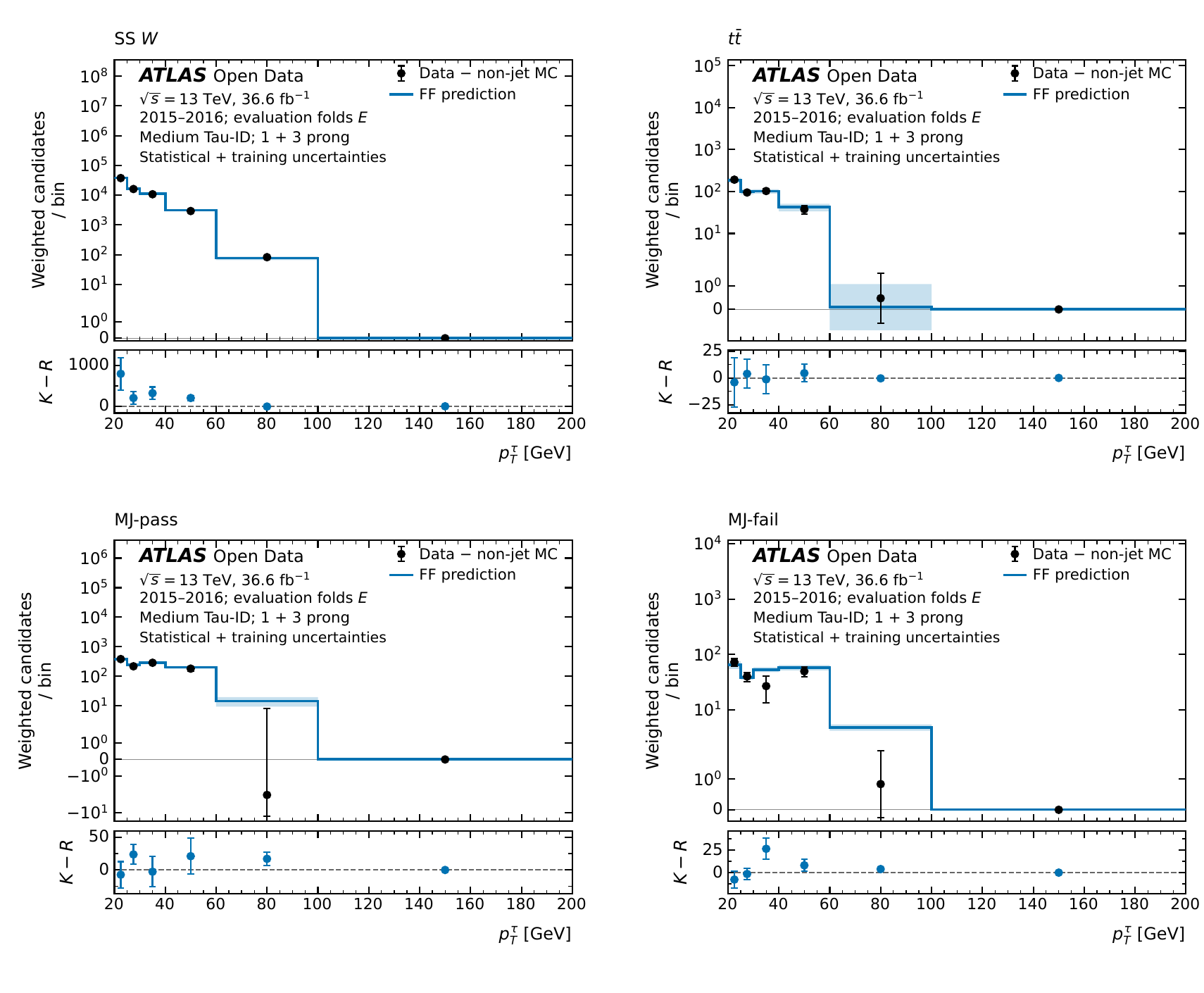}
\caption{Reference-FF checks for Medium identification versus tau $p_T$, combining prong
categories. Upper panels compare the reference-FF prediction $K$
(histogram and band) with ID data after non-jet-MC subtraction, $R$
(points). Lower panels show $K-R$, with uncertainties retaining the
correlation between $K$ and $R$. Bins follow their physical boundaries on a linear $p_T$ axis and show yields per
bin; symmetric-log $y$-axes in the upper panels retain signed yields. Bands and error bars show statistical and
training standard deviations from the same 20 repetitions as
Fig.~\ref{fig:reference-closure}}
\label{fig:reference-differential}
\end{figure*}

For Medium identification, the inclusive differences are $+2.26\%$ for SS $W$,
$+0.58\%$ for $t\bar t$, $+4.87\%$ for MJ-pass and $+15.51\%$ for
MJ-fail. At this working point, the trained FFs reproduce the inclusive
reference-region yields within approximately 16\%, supporting their
use in the CUFF combination. Figure~\ref{fig:reference-differential}
tests the prediction as a function of tau $p_T$.
Limited sample sizes lead to large statistical uncertainties in some
individual categories, reducing the precision of these checks.
These samples were excluded from the corresponding model training and
selection, but their checks informed method development.

\begin{table}[!htbp]
\centering\small
\caption{Reference-FF checks for Medium identification in evaluation folds $E$, combining
prong categories. The relative difference is $d=100(K_0-R_0)/R_0$, where $R_0$ is the ID data
yield after non-jet subtraction; $s_d$ is the statistical and training
uncertainty on $d$.}
\label{tab:source}
\begin{tabular*}{\columnwidth}{@{\extracolsep{\fill}}lrr@{}}
\toprule
Reference & $d$ [\%] & $s_d$ [pp] \\
\midrule
SS $W$ & +2.26 & 0.74 \\
$t\bar t$ & +0.58 & 7.47 \\
MJ-pass & +4.87 & 4.74 \\
MJ-fail & +15.51 & 7.24 \\
\bottomrule
\end{tabular*}
\end{table}

\subsection{Determination of the reference combination weights}
\label{sec:reference-combination}
\label{sec:mixture-checks}
The reference weights determine how the four FFs contribute to each
candidate's prediction. They depend on candidate and event variables
$v_i$ and on a summary $C_T$ of the target fail-Loose sample; the
subscript $T$ denotes the target region. The combination is
\begin{equation}
 \begin{gathered}
 \FF_{\cuff,i}(t)=\sum_{k=1}^{4}g_k(v_i,C_T)\FF_k(x_i;t),\\
 g_k\geq0,\quad\sum_k g_k=1.
 \end{gathered}
 \label{eq:mixture}
\end{equation}
The same weights are used at every identification threshold. The FF
inputs $x_i$ and combination inputs $v_i$ can differ, as specified in
Table~\ref{tab:features}.

Two networks determine $g_k$. They share the reference event pool and
data partitions with the FF networks. The first is a reference classifier trained
on data candidates in the four reference $\Azero$ samples, using the contamination weights
of Section~\ref{sec:reference-ff} and equal total training weight for
each reference. It has 16 inputs and two fully connected hidden layers
of 64 nodes with SiLU activation functions. Four softmax outputs,
$q_k^{(0)}(v)$, give the initial reference-class probabilities; the
superscript $(0)$ identifies this first step. The objective is the
weighted negative log-probability assigned to the candidate's known
reference label.

The second, the combination network, uses information about both the
candidate and the sample containing it. The sample summary has 14
entries: four averages of the classifier outputs and ten averages of
distinct pairwise products, including the four squares. Each average is
formed by subtracting the weighted non-jet-MC sum from the data sum and
dividing by the data yield after subtraction of the predicted non-jet
contribution. The same classifier is evaluated for data and MC.
The 37 inputs to the combination network are 19 candidate and event
variables, four classifier log-probabilities and the 14 summary entries.
Two fully connected SiLU layers with 64 nodes each give four corrections
to $\log q_k^{(0)}$. The corrected values are calibrated and transformed
by softmax to obtain nonnegative, unit-sum weights $g_k$.
The reference classifier is kept fixed while this network is trained.

Training uses reference $\Azero$ mixtures with varied, known proportions
$\boldsymbol\beta$. Within the reference training sample $T$, each reference is divided into two
halves: one supplies candidates and the other its sample summary.
The four summaries are combined with the chosen proportions to represent
the training mixture. Equal numbers of candidates are drawn from each
reference, with sampling probabilities proportional to their correction
weights $c_i$. The classification loss of a candidate from reference
$k$ is then multiplied by $4\beta_k$. Thus the contamination correction
enters through sampling and the desired mixture enters through the loss
weights. The objective remains the negative log-probability of the known
reference label; FF predictions do not enter this loss.

Both networks use AdamW and the reference $T/V/E$ assignments.
The corrected $\Azero$ candidates in $V$ are used to select the reference
classifier and calibrate its output probabilities. Fixed mixtures formed
from $V$ are used to select and calibrate the combination network. The precise objectives,
mixture sampling and probability calibration are specified in
Appendix~\ref{app:training}.

The combination is checked on eight mixtures of reference evaluation
samples with known anti-ID proportions. Each reference's signed
$\Azero$ yield is rescaled to the chosen fraction, and the same
coefficient weights its ID candidates. The coefficients remain fixed
across statistical replicas. The two sample halves exchange roles,
so each candidate's combination weights use a summary calculated from
the other half. The construction is specified fully in
Appendix~\ref{app:controlled-evidence}.

\begin{figure*}[tp]
\centering
\includegraphics[width=\linewidth]{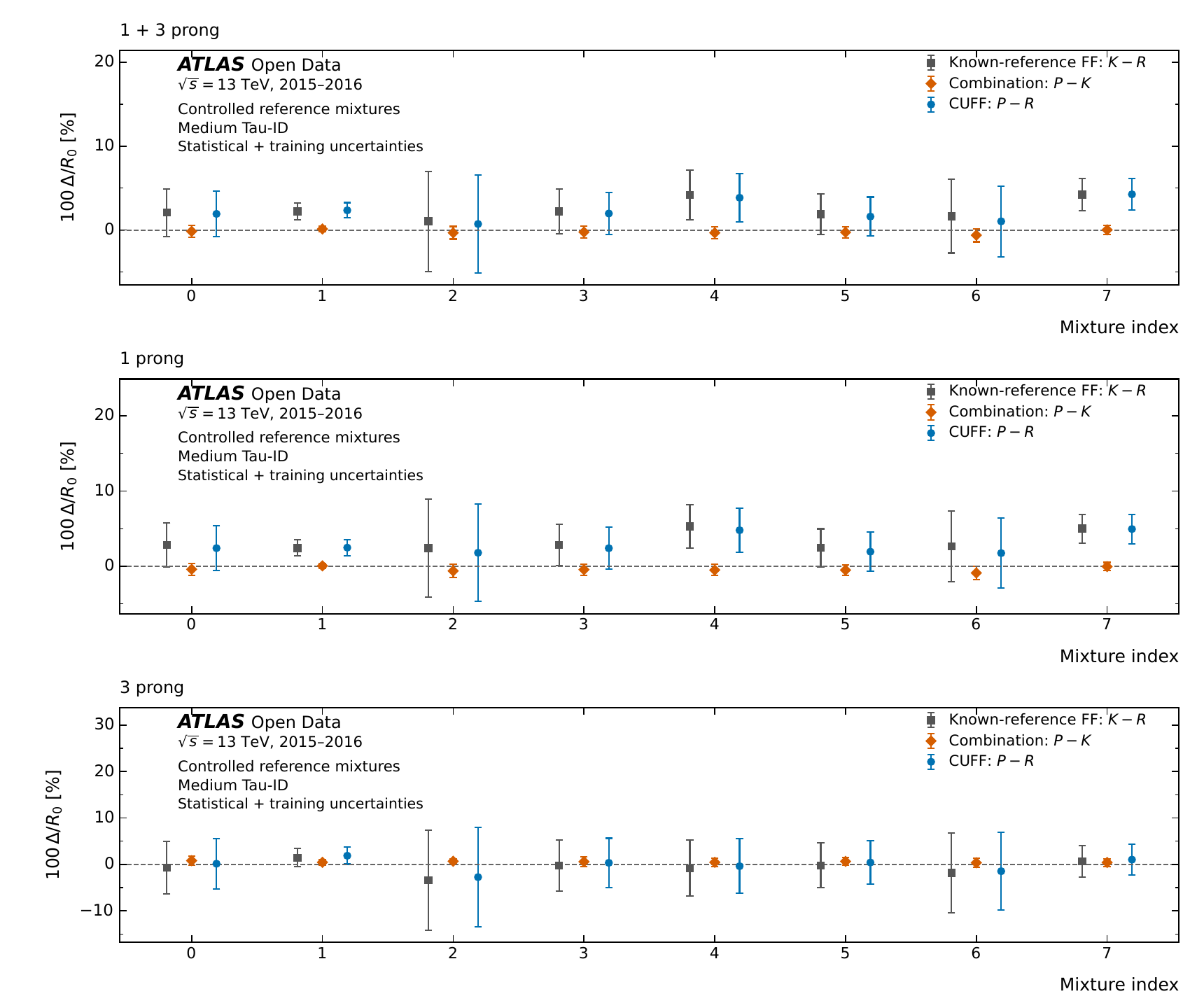}
\caption{Check of the learned reference combination for Medium identification in the
eight mixtures of Table~\ref{tab:mixtures}. $K$ applies each candidate's
known reference FF, $P$ uses the learned combination, and $R$ is the ID
data yield after non-jet-MC subtraction. The panels show the combined,
one-prong and three-prong samples from top to bottom.
Each compares the final prediction difference, the known-reference
difference and the change introduced by the combination. Error bars are standard deviations from the 20 paired
statistical and training replicas, multiplied by 100 and divided by fixed
$R_0$ (in percentage points).
The three differences are correlated}
\label{fig:controlled-decomposition}
\end{figure*}

For these mixtures, $K$ is the prediction obtained by applying the FF
associated with each candidate's known reference region, $P$ is the
prediction using the learned combination, and $R$ is the ID data yield
after non-jet-MC subtraction. The difference $P-K$ therefore measures the
change introduced by the learned weights for the existing FFs.
Figure~\ref{fig:controlled-decomposition} displays this change alongside
$K-R$ and the final difference $P-R$, all divided by the same nominal
reference $R_0$. Table~\ref{tab:mixtures} gives the inclusive results for Medium identification.

Across the eight mixtures, $100(P-R)/R_0$ ranges from $+0.72\%$ to
$+4.26\%$. The known-reference term $100(K-R)/R_0$ ranges from $+1.05\%$
to $+4.23\%$, while the additional combination term
$100(P-K)/R_0$ lies between $-0.62\%$ and $+0.13\%$.
The learned combination therefore reproduces the known-reference yield
response closely in these tests.

\begin{table*}[tp]
\centering\small
\caption{Inclusive prediction differences for Medium identification in mixtures with known
reference proportions. The entries of $\boldsymbol\beta$ follow the
order SS $W$, $t\bar t$, MJ-pass and MJ-fail. $d=100(P_0-R_0)/R_0$
and $s_d$ is its paired statistical and training uncertainty.}
\label{tab:mixtures}
\begin{tabular}{ccrr}
\toprule
Index & $\boldsymbol\beta$ & $d$ [\%] & $s_d$ [pp] \\
\midrule
0 & (0.25,0.25,0.25,0.25) & +1.93 & 2.71 \\
1 & (0.70,0.10,0.10,0.10) & +2.35 & 0.92 \\
2 & (0.10,0.70,0.10,0.10) & +0.72 & 5.82 \\
3 & (0.10,0.10,0.70,0.10) & +1.97 & 2.53 \\
4 & (0.10,0.10,0.10,0.70) & +3.86 & 2.89 \\
5 & (0.40,0.30,0.20,0.10) & +1.61 & 2.32 \\
6 & (0.05,0.15,0.75,0.05) & +1.04 & 4.22 \\
7 & (0.15,0.05,0.10,0.70) & +4.26 & 1.90 \\
\bottomrule
\end{tabular}
\end{table*}

\subsection{Application to a target region and variable Tau-ID thresholds}
\label{sec:application}
The application requires the target fail-Loose data candidates and an MC
estimate of the non-jet contribution in the same selection. For each
candidate, the trained reference networks evaluate four FFs and the
combination network determines their weights. Figure~\ref{fig:method-overview}
summarizes the calculation.

\begin{table}[!htbp]
\centering\small
\caption{Tau-ID score thresholds $t$, applied with the strict requirement $s>t$. For each prong category, $LM_f$ and $MT_f$ denote the fraction $f$ of the score interval from Loose to Medium or from Medium to Tight. All nine thresholds use the same fail-Loose $\Azero$ denominator.}
\label{tab:queries}
\begin{tabular*}{\columnwidth}{@{\extracolsep{\fill}}lrr@{}}
\toprule
Threshold & 1-prong cut & 3-prong cut \\
\midrule
Loose & 0.15 & 0.25 \\
$LM_{1/4}$ & 0.175 & 0.2875 \\
$LM_{1/2}$ & 0.2 & 0.325 \\
$LM_{3/4}$ & 0.225 & 0.3625 \\
Medium & 0.25 & 0.4 \\
$MT_{1/4}$ & 0.2875 & 0.4375 \\
$MT_{1/2}$ & 0.325 & 0.475 \\
$MT_{3/4}$ & 0.3625 & 0.5125 \\
Tight & 0.4 & 0.55 \\
\bottomrule
\end{tabular*}
\end{table}

The target $\Azero$ events are divided into five folds. To predict
candidates in one fold, the summary $C_T$ is calculated from the other
four folds using the corresponding trained reference classifier.
The combination network then assigns weights to the selected fold.
The five predictions are summed to obtain the target yield. The same
steps are applied to data and non-jet MC, evaluating each candidate's
FFs and combination weights from its own observables.

Let $D_{T,\Azero}$ and $M^{\rm non\text{-}jet}_{T,\Azero}$ denote the data and
non-jet MC candidates in the target application sample, and let $I(t)$
denote candidates passing the requested ID threshold. The predicted jet-to-tau yield
and its validation reference are
\begin{align}
 P(t)&=\sum_{i\in D_{T,\Azero}}\FF_{\cuff,i}(t)\nonumber\\*
       &\quad-\sum_{i\in M^{\rm non\text{-}jet}_{T,\Azero}}w_i\FF_{\cuff,i}(t),
       \label{eq:application}\\
 R(t)&=N^{\rm Data}_{T,I(t)}-N^{\rm non\text{-}jet,MC}_{T,I(t)}.
 \label{eq:reference}
\end{align}
Here $w_i$ is the signed MC event weight, including luminosity normalization,
and $N$ denotes a yield after selection. Subtracting the FF-weighted MC
removes its estimated contribution to the weighted $\Azero$ data yield.
The ID-region data enter the validation reference $R$, after the
prediction has been constructed.

\begin{figure*}[tp]
\centering
\includegraphics[width=\linewidth]{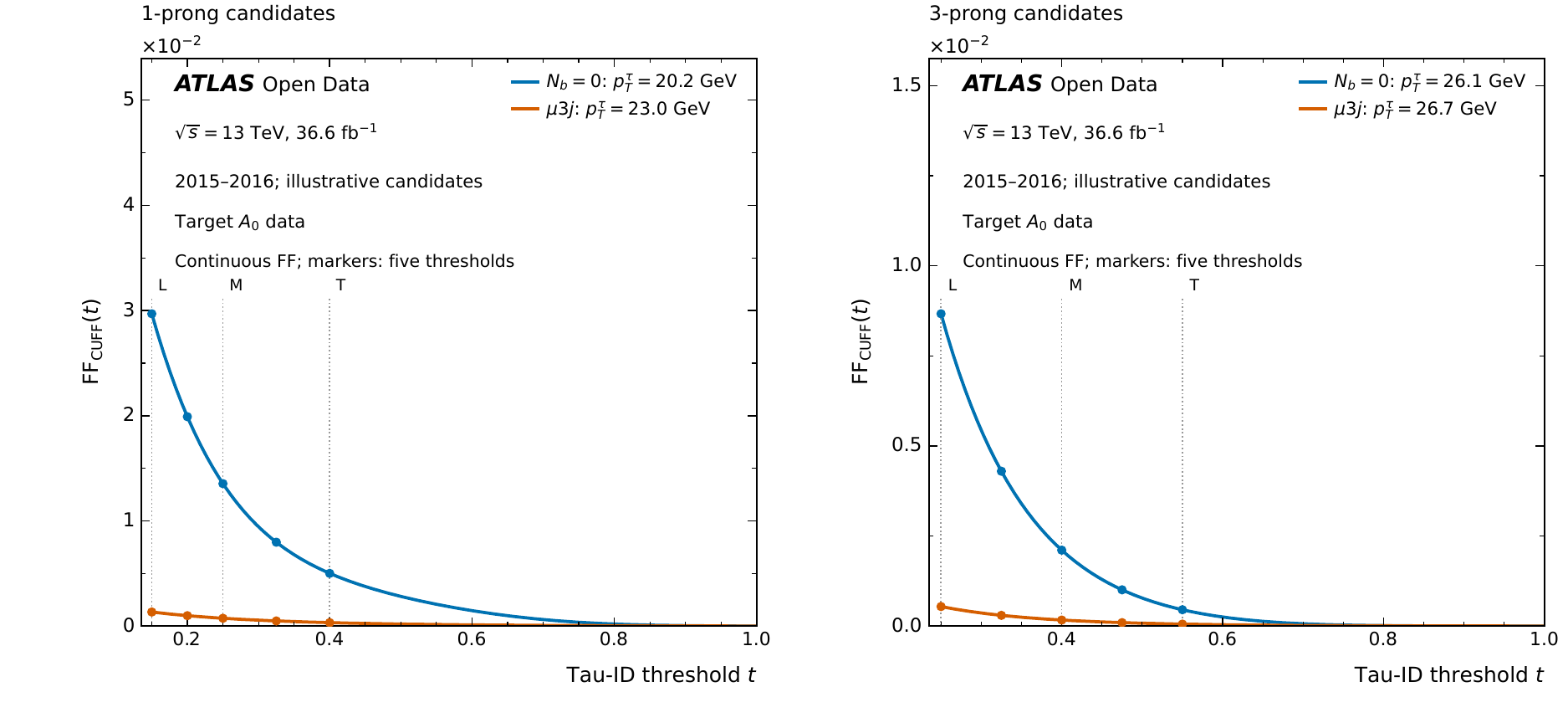}
\caption{Continuous CUFF responses for four example $\Azero$ data
candidates: a one-prong and a three-prong candidate from each validation
region. Candidates are selected by event identifier from the first
nominal fold, independently of their FF values or prediction agreement.
Curves use the trained FFs with fixed combination weights. Markers show
Loose, Medium, Tight and the two intervening score midpoints;
vertical guides mark the three working points (L, M, T).
These illustrate the model responses, rather than additional closure tests}
\label{fig:threshold-illustration}
\end{figure*}

Changing the threshold changes only the integral of the learned
passing-score distribution. The models, combination weights, sample
summary and $\Azero$ candidates remain the same.
Table~\ref{tab:queries} lists the nine thresholds used in this study:
the five defined in Section~\ref{sec:reference-ff}, plus thresholds at the
quarter and three-quarter positions of the Loose--Medium and
Medium--Tight score intervals.
Figure~\ref{fig:threshold-illustration} illustrates the dependence of
$\FF_{\cuff}$ on the Tau-ID threshold for four example data candidates
failing Loose identification: a one-prong and a three-prong candidate
from each of the $N_b=0$ and $\mu3j$ validation regions.

The comparisons with ID data at all nine thresholds are presented in
Section~\ref{sec:nine-thresholds}.

\section{Validation and results}
\label{sec:validation-strategy}
\label{sec:results}
In this study, we compare the yields and kinematic distributions
predicted by CUFF with those obtained from data after non-jet-MC
subtraction in the two validation regions. We evaluate these predictions
at different Tau-ID thresholds. For Loose identification, we also compare CUFF with UFF adapted to the available
public inputs.
Using $P$ and $R$ from Eqs.~\eqref{eq:application} and~\eqref{eq:reference},
with nominal values $P_0$ and $R_0$, we define the
relative difference as $d=100(P_0-R_0)/R_0$, expressed in percent, and
denote its statistical and training uncertainty by $s_d$, expressed in
percentage points (pp). The error bars correspond to one standard
deviation and include statistical and training uncertainties, with
correlations between the prediction and reference taken into account. Systematic
uncertainties are estimated separately in Section~\ref{sec:uncertainties}.

The SS $W$ reference was used during development. For the initial OS
validation, the selections, models, target-summary definition and the five
thresholds of Section~\ref{sec:reference-ff} were fixed before the ID results were examined;
the $\Azero$ predictions were constructed first. Later diagnostics and
the four additional thresholds in Table~\ref{tab:queries} reuse these events and the fixed models.

\subsection{Predictions in the validation regions}
\label{sec:OS}
The CUFF prediction is tested in the mutually exclusive $N_b=0$ and
$\mu3j$ validation regions. Both require opposite-sign muon--tau pairs,
with different jet selections as defined in
Table~\ref{tab:regions}. The total CUFF predictions and data references
at the three standard working points are listed in
Appendix~\ref{app:validation-tables}. The inclusive
predictions lie below the reference in both regions. For Medium identification, the
relative differences are $-21.85\%$ in $N_b=0$ and $-45.84\%$ in
$\mu3j$, with statistical and training uncertainties of 1.58 and
6.03 pp, respectively.

\begin{figure*}[tp]
\centering
\includegraphics[width=\linewidth]{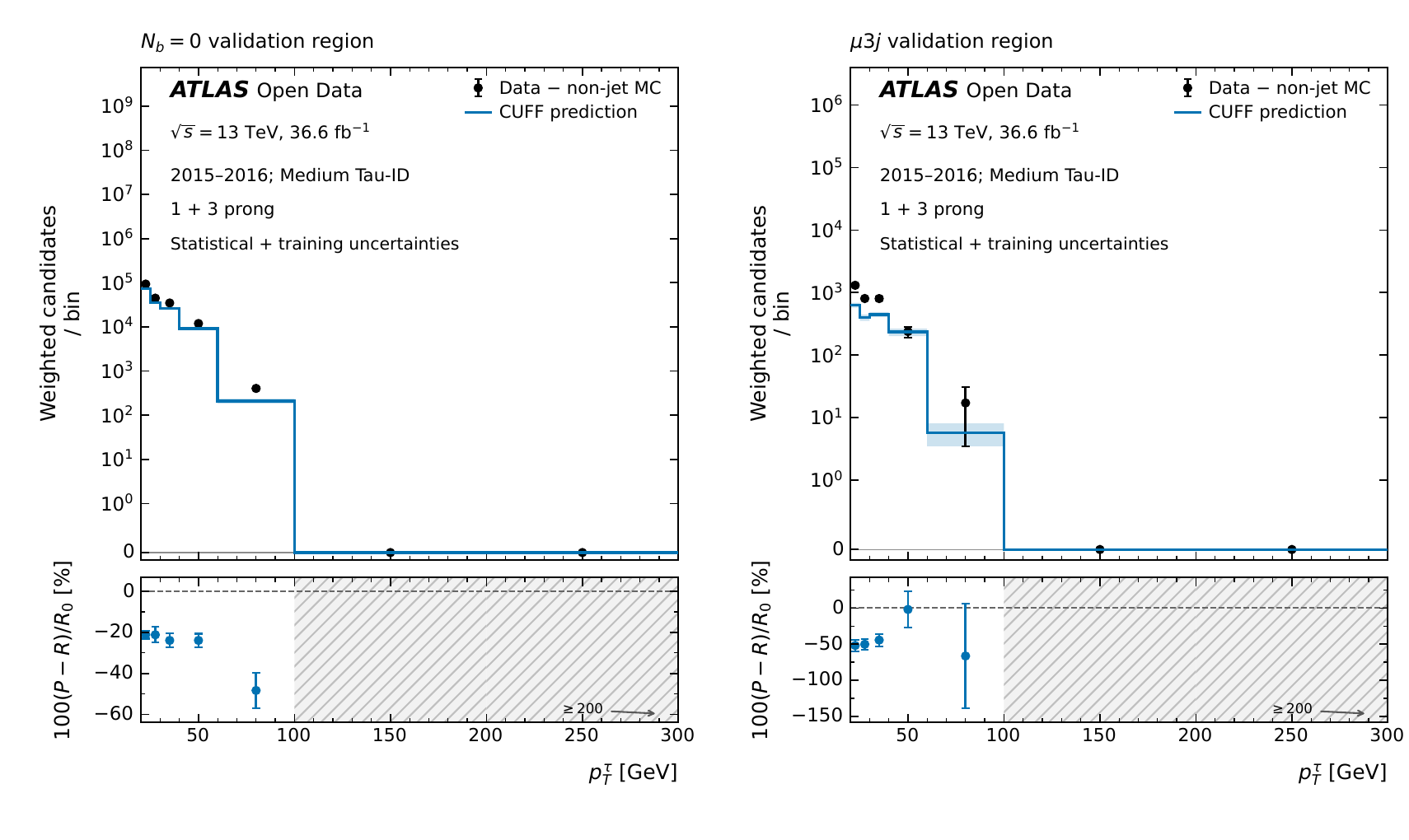}
\caption{Tau-$p_T$ distributions for Medium identification in the two validation regions,
combining prong categories. Upper panels compare CUFF (histogram and
band) with ID data after non-jet-MC subtraction (points); lower panels
show their relative difference. Bands and error bars are standard
deviations from 20 resampling and training repetitions, with
prediction--reference correlations retained in the lower panels.
The $y$-axes in the upper panels use symmetric-log scales. Finite bins follow their physical
boundaries; the arrow marks the final display interval containing the
$p_T\geq200$ GeV overflow yield. Bins with $R_0\leq0$ in the lower panels are hatched.
Systematic uncertainties are estimated separately in
Section~\ref{sec:uncertainties}}
\label{fig:target-medium}
\end{figure*}

Figure~\ref{fig:target-medium} compares the predictions and data
references for Medium identification as a function of tau $p_T$. The predictions follow the overall
decrease of the spectra but underestimate the yields in the 20--40 GeV
range in both regions. The UFF validation also reported region-dependent
shifts of the central predictions within the quoted uncertainties~\cite{UFF}.
The deviations in the 20--40 GeV range are encompassed by the MC transfer
variations estimated in Section~\ref{sec:transfer-results}.
The remaining underprediction may reflect differences in the conditional
fake rates between reference and validation regions and residual
mismodelling. Further validation with larger reference samples and targeted
checks of these dependencies can guide improvements.

\subsection{Validation of CUFF across Tau-ID thresholds}
\label{sec:nine-thresholds}
\label{sec:nine}
We validate CUFF at nine Tau-ID thresholds: Loose, Medium and Tight,
together with six intermediate thresholds between these working points.
Table~\ref{tab:queries} gives the numerical thresholds for both prong categories.

\begin{figure*}[tp]
\centering
\includegraphics[width=\linewidth]{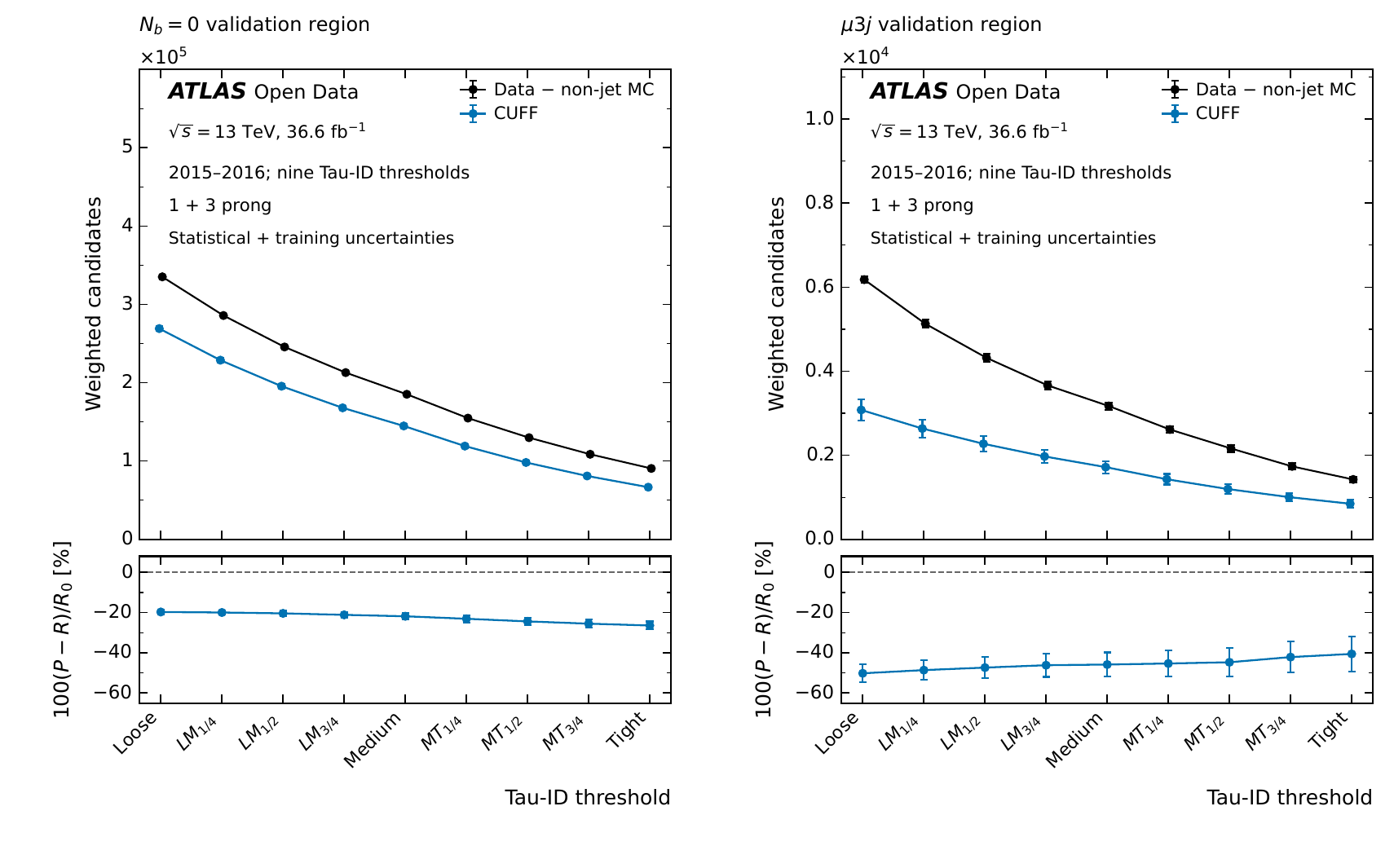}
\caption{Application of the same trained CUFF models at nine identification
thresholds in the $N_b=0$ (left) and $\mu3j$ (right) validation regions,
summing one-prong and three-prong candidates. The upper panels compare the
prediction $P$ with the data reference $R$ after non-jet subtraction; their
error bars are the respective standard deviations across the resampling
and training repetitions.
The lower panels show $100(P_0-R_0)/R_0$, with the correlated difference
uncertainty evaluated at fixed nominal $R_0$. Only statistical and training
uncertainties are displayed. The relative differences change
gradually as the identification threshold is tightened}
\label{fig:nine}
\end{figure*}

Figure~\ref{fig:nine} shows that the relative
yield differences vary gradually from $-19.71\%$ at Loose to $-26.43\%$
at Tight in $N_b=0$, and from $-50.19\%$ to $-40.56\%$ in $\mu3j$.
The intermediate thresholds follow the same trend.
The numerical results at all nine thresholds are given in
Appendix~\ref{app:validation-tables}.

These results demonstrate consistent CUFF background estimation as the
Tau-ID threshold is varied. CUFF provides fake factors at arbitrary
thresholds within the modelled score range without additional
threshold-specific measurements. In contrast, the published UFF
implementation requires corresponding fake factors and template inputs
for each identification working point~\cite{UFF}.

\subsection{Comparison with UFF adapted to public data}
\label{sec:UFFmethod}
\label{sec:UFFresult}
For Loose identification, CUFF is compared with UFF adapted to the available public
inputs~\cite{UFF}. Both methods use the same four reference regions,
event selections, data partitions, non-jet estimates and statistical
replicas. The adapted reference FFs are measured directly from corrected
ID and anti-ID yields in tau-$p_T$/prong bins, with $p_T$ boundaries at
20, 25, 30, 40 and 60 GeV and an open final bin.

\begin{figure*}[tp]
\centering
\includegraphics[width=\linewidth]{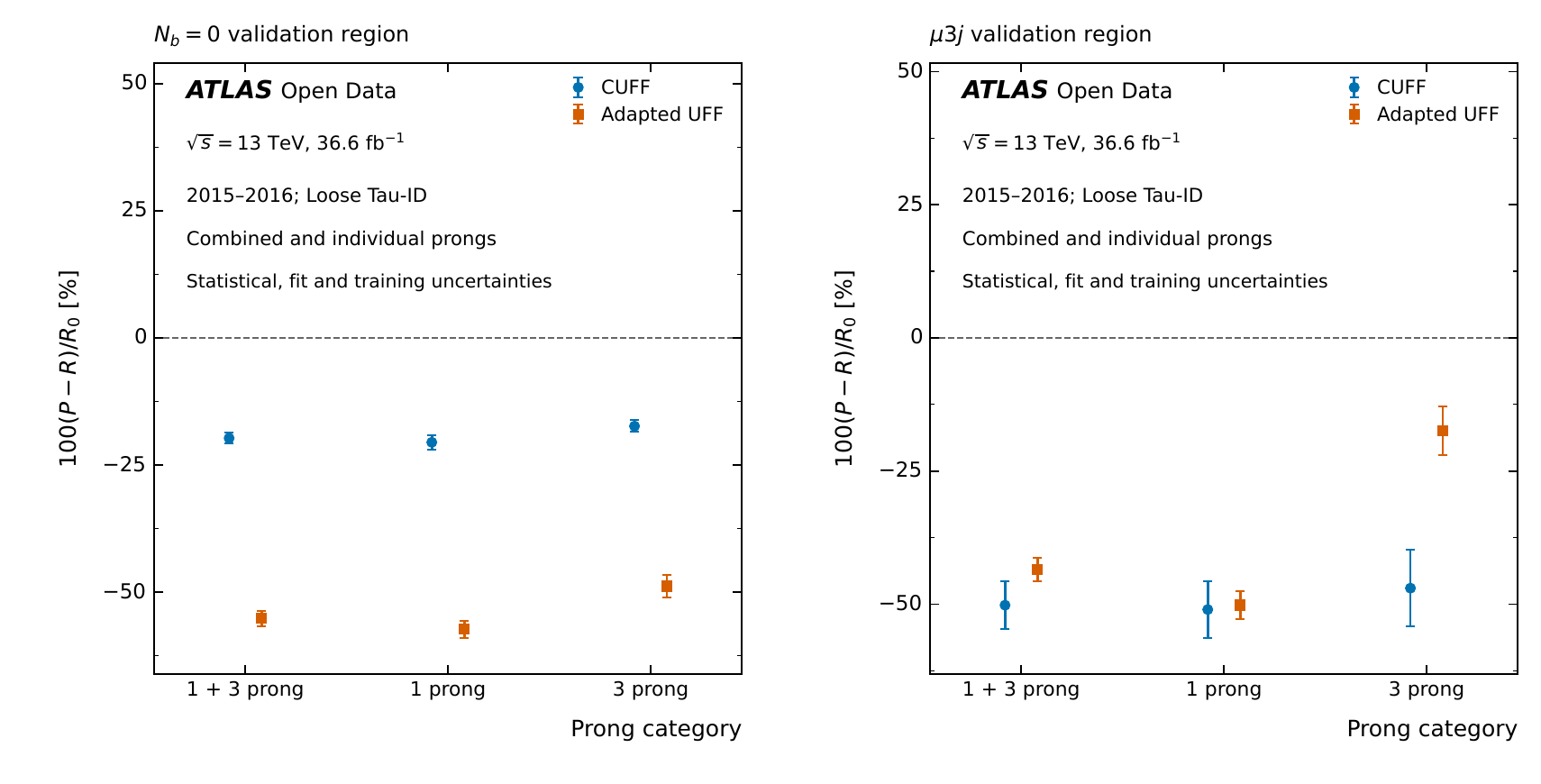}
\caption{Loose-ID relative prediction differences for CUFF and adapted
UFF, for the combined and individual prong samples in each validation
region. The reference is ID data after non-jet-MC subtraction.
Error bars are paired prediction-minus-reference standard deviations
relative to fixed nominal $R_0$, including statistical, template-fit
and training fluctuations}
\label{fig:UFFcomparison}
\end{figure*}

The associated-jet track multiplicity replaces the published UFF q/g
discriminant. Its reference distributions are fitted to the target
$\Azero$ distribution using iterated weighted least squares, including
finite-template uncertainties and the target covariance. The coefficients
sum to one and may be negative. The prediction uses the same linear
combination as Eq.~\eqref{eq:intro-uff}. In tau-$p_T$/prong categories
with insufficient statistics, the template fit uses samples combined
over $p_T$ within the same prong category, or over both prong categories
if necessary. Details of this public-input implementation are retained in
Appendix~\ref{app:comparison-details}.

Figure~\ref{fig:UFFcomparison} shows the relative differences from the
data reference, with numerical values listed in
Appendix~\ref{app:validation-tables}. In $N_b=0$, they are
$-19.71\%$ for CUFF and $-55.14\%$ for adapted UFF; in $\mu3j$, they
are $-50.19\%$ and $-43.52\%$, respectively.
The inclusive adapted-UFF-minus-CUFF difference in $\mu3j$ is
$+6.67\pm5.28$ pp relative to $R_0$. The uncertainty accounts for data
and MC statistics, template-fit and training fluctuations, and the
correlation between the two predictions. Compared with adapted UFF,
CUFF gives closer agreement with the data reference in $N_b=0$ and a
comparable inclusive prediction in $\mu3j$.

\subsection{Multivariable shape comparison}
\label{sec:shape-strategy}
\label{sec:shape-results}
For Loose identification, we compare CUFF and adapted UFF in seven kinematic observables,
including variables used as model inputs and variables not supplied to
the models. To compare shapes independently of the total-yield
normalization, each prediction and data reference is divided by its own
total in the same region and prong category:
\begin{equation}
 \begin{aligned}
 \widehat P_i&=\frac{P_i}{\sum_j P_j},\qquad
 \widehat R_i=\frac{R_i}{\sum_j R_j},\\
 D_{\rm shape}&=50\sum_i\left|\widehat P_i-\widehat R_i\right|.
 \end{aligned}
 \label{eq:shape-distance}
\end{equation}
Here $i$ labels a bin, and smaller $D_{\rm shape}$ indicates closer
normalized shapes. The factor $50=100/2$ converts to percentage points
and avoids counting a bin excess and its compensating deficit twice.

\begin{figure*}[tp]
\centering
\includegraphics[width=\linewidth]{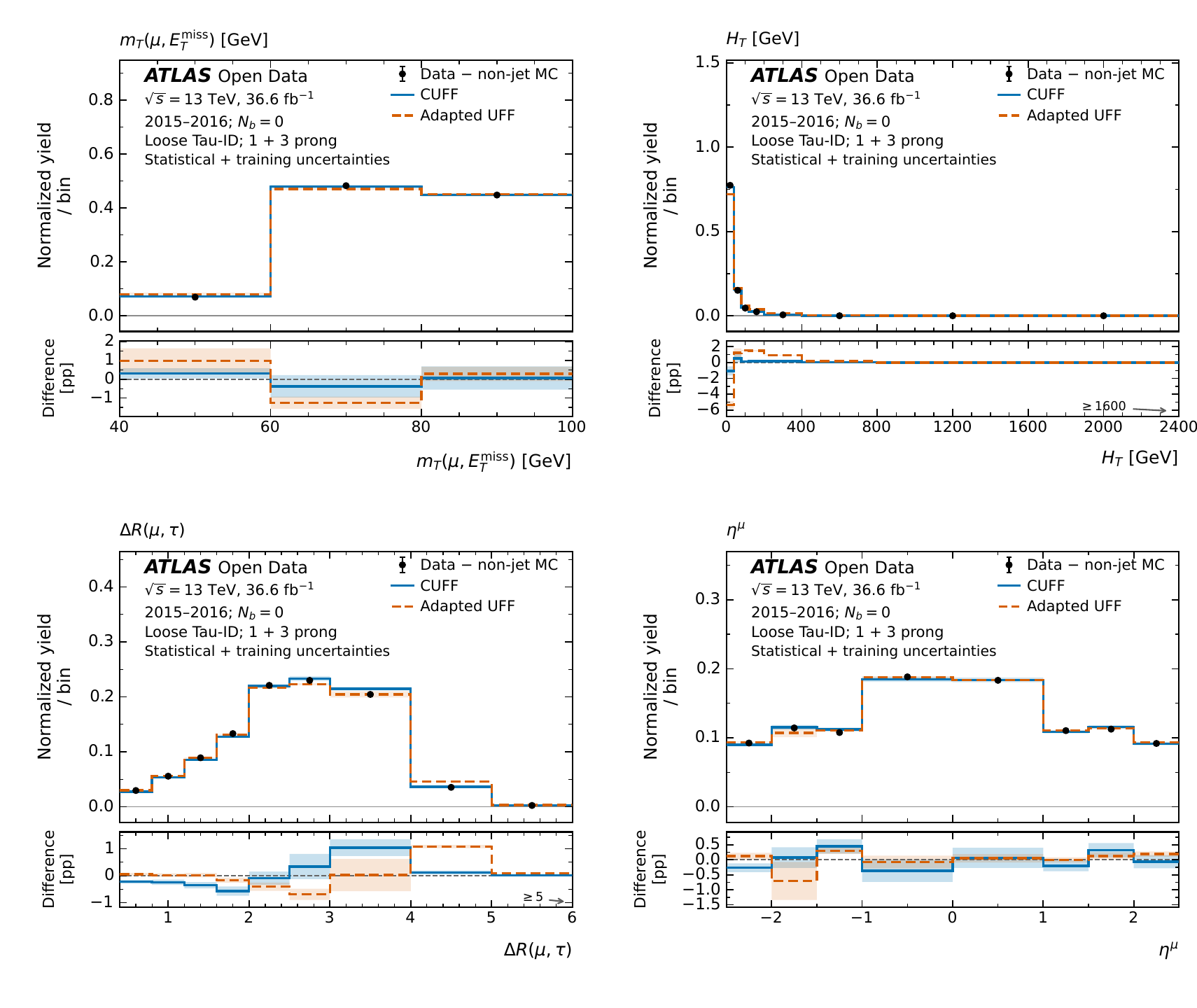}
\caption{Normalized Loose-ID distributions in the $N_b=0$ validation region,
combining prongs, for four observables not explicitly supplied to the
models. The upper panels show CUFF and adapted UFF as step histograms,
with the ID data after non-jet subtraction shown as points. Each
distribution is normalized by its own total. Shaded model bands show
statistical and training uncertainties; data error bars show statistical
uncertainties.
The lower panels show $100(\widehat P_i-\widehat R_i)$ in percentage
points, with uncertainties including the correlation between prediction
and reference. Signed bins are retained. Finite bins are drawn at their
physical boundaries; nonempty overflow bins are shown in the final
display intervals and marked by arrows. Bins that are exactly zero in the
nominal result and all repetitions are omitted}
\label{fig:shape-nb0}
\end{figure*}

\begin{figure*}[tp]
\centering
\includegraphics[width=\linewidth]{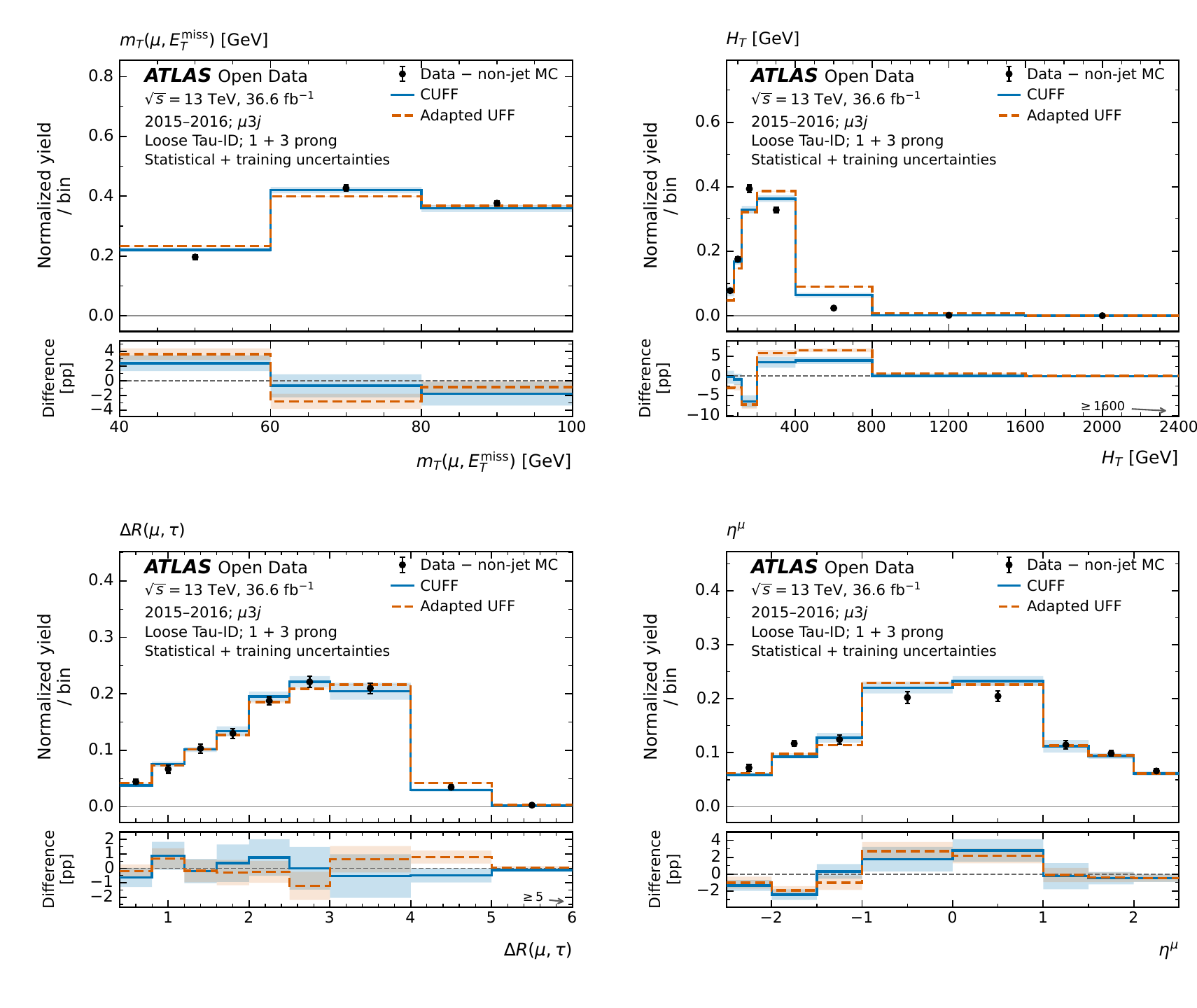}
\caption{The same normalized-shape comparison as in
Fig.~\ref{fig:shape-nb0}, for the $\mu3j$ validation region.
The error bars and bands include statistical and training effects;
systematic uncertainties are not included}
\label{fig:shape-mu3j}
\end{figure*}

The numerical shape comparisons are summarized in
Appendix~\ref{app:validation-tables}.
CUFF gives a smaller shape distance for five of the seven observables
in each validation region. These correlated comparisons show an overall
improvement across several distributions, with the size of the gain
depending on the observable and region.
For the model inputs---tau $p_T$, $E_T^{\rm miss}$ and selected-jet
multiplicity---the improvements are clearest in $N_b=0$;
the tau-$p_T$ shape is also improved in $\mu3j$.

The other four observables are $m_T(\mu,E_T^{\rm miss})$,
$H_T$ (the scalar sum of selected-jet transverse momenta),
$\Delta R(\mu,\tau)$ and muon $\eta$.
They are not explicit inputs to the FF, combination or contamination
models, or to the target summary. Their distributions test the
generalization of the learned reweighting to correlated observables
that were not supplied as inputs.
Figures~\ref{fig:shape-nb0} and~\ref{fig:shape-mu3j} show these comparisons
in the two validation regions. The improvement extends to $H_T$ in both
regions: the reductions in $D_{\rm shape}$ relative to adapted UFF are
$4.30\pm0.76$ pp in $N_b=0$ and $5.57\pm1.49$ pp in $\mu3j$.
The angular distributions show no consistent preference
between the methods. The gains in both input and noninput distributions
support the ability of CUFF to describe several correlated observables
with a common background prediction, and demonstrate a practical
advantage over adapted UFF for the distributions improved in this study.

\section{Uncertainty estimation}
\label{sec:uncertainties}
This section presents initial estimates of the uncertainties in CUFF
predictions. The sources considered are finite data and MC statistics,
model training, modelling of the reference FFs and combination weights,
transfer between regions, and non-jet normalization and modelling.

The nominal prediction $P_0$ is the FF-weighted target $\Azero$ data
yield after subtraction of the correspondingly weighted non-jet MC.
The reference $R_0$ is the ID data yield after non-jet-MC subtraction.
We use the relative difference $d$ and its statistical and training
uncertainty $s_d$ introduced in Section~\ref{sec:results}.
The uncertainty contributions are estimated
from statistical resampling, reference-region checks and comparisons
between simulated regions, as detailed below.

For a systematic uncertainty, we change the relevant inputs or model
components according to the prescription described below and recalculate
the affected yields. Each such alternative calculation is called a
systematic variation.

Table~\ref{tab:uncertainty-components} summarizes the statistical, training
and jet-fake modelling uncertainties for the Medium identification selection.
Its statistical and training entry gives $s_d$. For each
systematic variation, let $P_{\rm varied}$ and $R_{\rm varied}$ denote
the recalculated prediction and reference. The reference remains $R_0$
for variations that do not affect the ID subtraction.
The table reports the range
of shifts in their difference,
$100[(P_{\rm varied}-R_{\rm varied})-(P_0-R_0)]/R_0$,
over the specified variations.

\begin{table*}[tp]
\centering\small
\caption{Estimated statistical, training and jet-fake modelling uncertainties in the inclusive comparison for Medium identification. All entries are in percentage points relative to the nominal ID reference $R_0$. The statistical and training entry is one standard deviation. The other entries give the smallest and largest shifts in $P-R$ among the specified systematic variations. The separate FF and weight entries are included in the combined entry and must not be added to it.}
\label{tab:uncertainty-components}
\begin{tabular}{llrr}
\toprule
Source of uncertainty & Type & $N_b=0$ [pp] & $\mu3j$ [pp] \\
\midrule
Statistical and training & Standard deviation & $1.58$ & $6.03$ \\
Reference FF modelling alone & Range of shifts & $[-2.43, +2.43]$ & $[-2.59, +3.11]$ \\
Reference-weight modelling alone & Range of shifts & $[-0.91, +0.91]$ & $[-0.87, +0.87]$ \\
Combined FF and weight modelling & Range of shifts & $[-2.96, +4.12]$ & $[-3.40, +3.89]$ \\
MC transfer between regions & Range of shifts & $[-69.57, +57.55]$ & $[-49.89, +57.57]$ \\
Low-NNJvt data control & Range of shifts & $[+0.0000, +0.0013]$ & $[+0.0000, +0.0202]$ \\
\bottomrule
\end{tabular}
\end{table*}

\subsection{Statistical fluctuations and model training}
\label{sec:statistical-uncertainties}
To estimate how finite sample sizes and model training affect the
prediction, the training and prediction are repeated using statistically
resampled data and MC samples. For each of twenty repetitions, each data
or MC event is assigned a random number of copies, drawn from a Poisson
distribution with mean one. An event can therefore be omitted or
contribute more than once. This number, called the multiplicity,
multiplies its event weight.
The same multiplicity is used wherever the event contributes to training,
prediction or the ID reference. The relevant models are retrained and
the target summary is recalculated in each repetition.

The index $r=1,\ldots,20$ labels these repetitions, with $P_r$ and $R_r$
denoting the prediction and data reference calculated in the same
repetition. The five evaluation-fold predictions are summed in each
repetition. The relative difference and its statistical and training
uncertainty are calculated as
\begin{equation}
 \begin{aligned}
 d(t)&=100\frac{P_0(t)-R_0(t)}{R_0(t)},\\
 s_d(t)&=100\frac{\operatorname{SD}_{r=1,\ldots,20}[P_r(t)-R_r(t)]}{R_0(t)}.
 \end{aligned}
 \label{eq:residual}
\end{equation}
Here SD denotes the sample standard deviation across the twenty
repetitions. Calculating
$P_r-R_r$ within each repetition preserves the relevant event
correlations. This procedure includes statistical fluctuations and
model-training variation, so no additional statistical term is added.
For Medium identification, $s_d$ is 1.58 pp in $N_b=0$ and 6.03 pp in $\mu3j$.

The mean of $P_r-R_r$ across the repetitions can differ from the nominal
difference $P_0-R_0$. These shifts are reported separately, rather than
used to correct the nominal difference. For Medium identification, the inclusive shifts relative to $R_0$ in the eight
reference mixtures range from 1.98 to 6.56 pp. Details of event grouping, covariance calculation and
training differences between repetitions are given in
Appendix~\ref{app:statistical-details}.

To test the statistical uncertainty estimates, we generate 100
independent synthetic samples with
known distributions, referred to as pseudo-experiments. They use two
fixed mixtures of the four reference sources, labelled $A$ and $B$;
their proportions are given in Appendix~\ref{app:statistical-details}.
In each pseudo-experiment, the FF and combination models are trained
and the uncertainty is estimated using twenty resampling repetitions.
A nominal 95\% interval is formed around $P-R$ using approximately
1.96 times its estimated standard deviation. We count how often this
interval contains zero, corresponding to equality of the prediction
and the finite-sample ID reference; this fraction is the coverage.
For the inclusive yields, it is 93--97\%
(Table~\ref{tab:learned-toy-coverage}). Coverage is lower in some
categories with limited statistics. Additional checks show that an
interval describing statistical fluctuations need not cover a
persistent bias in the trained prediction. The detailed tests and
individual results are given in
Appendix~\ref{app:statistical-details}.

\begin{table}[!htbp]
\centering
\caption{Coverage of nominal 95\% intervals for $P-R$, combining prongs:
entries count intervals containing zero out of 100 independent
pseudo-experiments. $P$ is the FF prediction and $R$ the finite-sample
ID reference. Mixtures $A$ and $B$ are defined in
Appendix~\ref{app:statistical-details}; target sizes are expected
fake-$\Azero$ counts. Results across thresholds and target sizes are correlated.}
\label{tab:learned-toy-coverage}
\setlength{\tabcolsep}{3pt}
\begin{tabular*}{\columnwidth}{@{\extracolsep{\fill}}lrrrr@{}}
\toprule
Mixture & \shortstack{Expected\\fake $\Azero$} & Loose & Medium & Tight\\
\midrule
$A$ & 600  & 93 & 96 & 95\\
$A$ & 6000 & 97 & 93 & 96\\
$B$ & 600  & 93 & 93 & 95\\
$B$ & 6000 & 97 & 96 & 96\\
\bottomrule
\end{tabular*}
\end{table}

\subsection{FF and reference-weight modelling}
Differences between predictions and comparison yields in the
reference-FF and controlled-mixture checks are used to set the size of
systematic variations associated with learning the FFs and their
combination. The reference evaluation candidates are
divided into disjoint subsets: one determines the variations and the
other checks their effect. Candidates are grouped by prong, NNJvt, kinematics
and non-overlapping Tau-ID score intervals. The intervals are bounded by
the five thresholds of Section~\ref{sec:reference-ff} and the upper
score limit of one. The statistical precision requirements and
grouping rules are given in Appendix~\ref{app:learning-response}.

For the FF variation, $K$ denotes the yield predicted using the FF of
the candidate's known reference region, and $R$ is the ID data yield
after non-jet subtraction in the same group. In groups meeting the
statistical requirements, $b_F=R/K-1$ gives the fractional change
needed to bring $K$ to $R$. This measured difference sets two opposite
shifts of the FF contribution in that score interval. With
$\theta\in\{-1,0,+1\}$ labelling the choices, the multiplier is
$\max(0,1+\theta b_F)$. Zero selects the unchanged FF, and the two
other choices apply the measured difference in opposite directions.
The maximum keeps the multiplier nonnegative.

Changing the reference FFs can also change the discrepancy attributed
to their combination. We therefore evaluate the combination in controlled
mixtures with known reference proportions for each specified set of FF
changes, labelled $h$. Let $K(h)$ be the prediction using the known
reference assignments and $P(h)$ the prediction using the learned
combination, both evaluated with the same changed FFs. The fractional
change that brings $P(h)$ to $K(h)$ is
\begin{equation}
 b_g(h)=\frac{K(h)}{P(h)}-1.
 \label{eq:combination-response}
\end{equation}
The combination variation uses the same three multiplier choices
constructed from $b_g(h)$, with its direction chosen separately from
the FF variation. These nonnegative multipliers modify the combined
prediction in each score interval. The trained combination network is
unchanged; the variation modifies its predicted yield contribution.
The comparison is repeated for each $h$, so the combined FF and weight
variation includes the dependence of $b_g$ on the changed FFs.

The isolated FF and reference-weight rows in
Table~\ref{tab:uncertainty-components} show their separate effects.
The combined ranges for Medium identification are
$[-2.96,+4.12]$ pp in $N_b=0$ and $[-3.40,+3.89]$ pp in $\mu3j$.
The specified combinations of changes and their construction are given
in Appendix~\ref{app:learning-response}.

\subsection{Transfer between reference and validation regions}
\label{sec:transfer-results}
This uncertainty estimates the stability of the trained reference FF
models when applied in different event environments. Following
the UFF approach, we compare regions in MC using tau candidates
identified by truth information as jet fakes~\cite{UFF}. The comparison
is made separately for each simulated jet type. It is sensitive to
changes in the fake rates and to residual modelling differences as
the candidate distributions change. These candidates are
distinct from the genuine-tau and lepton contributions used for non-jet
subtraction.

\begin{figure*}[tp]
\centering
\includegraphics[width=\linewidth]{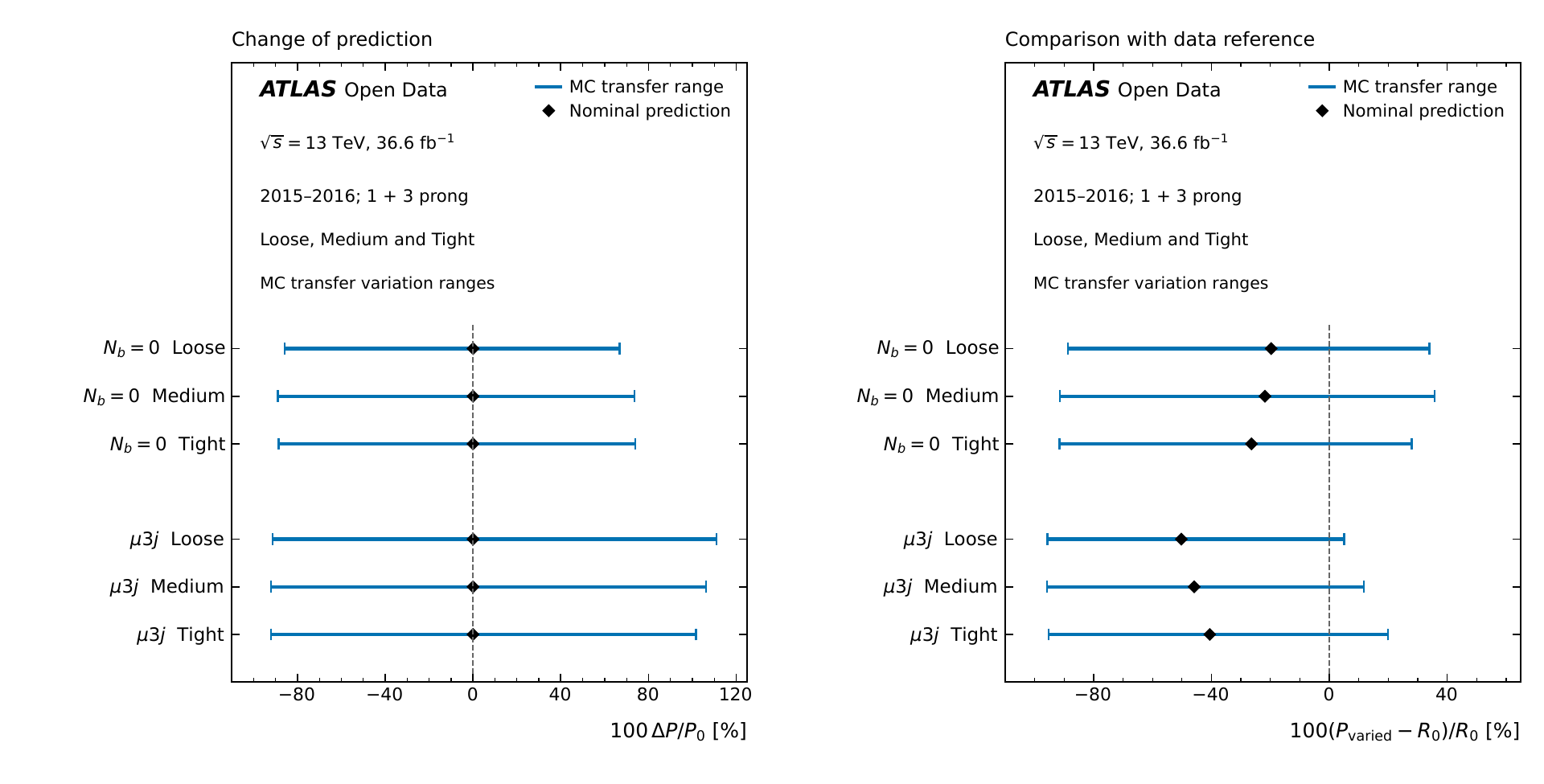}
\caption{Ranges obtained from the MC transfer variations at Loose,
Medium and Tight.
The left panel gives the change relative to the nominal prediction
$P_0$; the right gives the resulting prediction-minus-reference
comparison relative to fixed $R_0$, for which zero denotes equality.
The panels show the effect of the same MC transfer variations with different denominators.
Only the MC transfer variations are applied. The ranges retain the
correlations specified in the text and are not confidence intervals}
\label{fig:migration}
\end{figure*}

For FF model $k$, let $a$ be its reference region and $e$ a comparison
region. The same trained model evaluates an FF for each simulated
jet-fake $\Azero$ candidate using its input variables in the two regions.
Within a given jet-type, prong, kinematic
and score-interval group, let $I$ denote the simulated ID yield and
$P_k$ the corresponding prediction from FF $k$. The ratio $I/P_k$
measures their agreement in one region. Comparing this ratio between
the two regions gives
\begin{equation}
 D_{k,e}=\frac{I_e/P_{k,e}}{I_a/P_{k,a}},
 \label{eq:migration}
\end{equation}
Comparisons must meet requirements on event statistics and ratio
precision. Neighbouring groups are combined when necessary to obtain
a stable estimate. For each category that meets these requirements,
the accepted $D_{k,e}$ values are used to define lower and upper FF
variations.

The changes are propagated separately for the four reference FFs and
two prong categories, giving eight contributions. Within each such
contribution, the lower or upper choice is applied across its kinematic
and score categories; the magnitude of the FF variation can differ
between categories. The FF variations are propagated to both the
weighted data yield and the weighted non-jet-MC subtraction, keeping
the combination weights fixed. The statistical
requirements and construction of the final range are specified in
Appendix~\ref{app:migration-algorithm}.
For the MJ-fail FF component applied to candidates failing NNJvt,
differences measured in a data control sample selected at low NNJvt
replace the corresponding MC estimate. This uncertainty is listed
separately in Table~\ref{tab:uncertainty-components}.

The largest estimated source of uncertainty is the transfer between regions.
For Medium identification, it changes the prediction relative to $P_0$ by
$[-89.01,+73.63]\%$ in $N_b=0$ and $[-92.12,+106.29]\%$ in $\mu3j$.
Figure~\ref{fig:migration} displays these ranges at the standard working
points and their effect on the prediction-minus-reference comparison.
Differences between environments, limited MC sample sizes and taking
the smallest and largest accepted ratios all contribute to their width.
Some influential $\mu3j$ groups contain only eight or nine ID MC records.

Relative to the data reference, these same variations for Medium identification give differences
$100(P_{\rm varied}-R_0)/R_0$ of approximately $[-91.4,+35.7]\%$ in
$N_b=0$ and $[-95.7,+11.7]\%$ in $\mu3j$.

\subsection{Summary and scope of the uncertainty estimation}
\label{sec:uncertainty-scope}
In addition to the uncertainties discussed above, we examined the
non-jet MC subtraction using luminosity and $t\bar t$ cross-section
variations and a representative study of $t\bar t$ scale and radiation
variations. The latter change the inclusive prediction for Medium identification by
$[-0.074,+0.037]\%$ in $N_b=0$ and $[-0.352,+2.617]\%$ in $\mu3j$,
relative to $P_0$. The tested non-jet effects are smaller than the
FF variations between regions, suggesting a subdominant contribution
in this proof of concept. A more complete study of other processes is
left to future applications in physics analyses.

The dominant estimated source of uncertainty is the transfer between regions.
The statistical and training uncertainties and the combined FF and
weight variations have smaller effects on the inclusive results for Medium identification.
The first row of Table~\ref{tab:uncertainty-components} is a standard
deviation. The remaining rows give the smallest and largest shifts
among the specified systematic variations. Some rows share sources of uncertainty:
in particular, the separate FF and weight entries are already included
in their combined variation. These ranges must therefore not be added
in quadrature as independent Gaussian uncertainties.

For categories in which the available samples do not allow a reliable
variation to be determined, the contribution is left unchanged in the
displayed calculation. Its uncertainty has not been estimated.

Statistical and training uncertainties are evaluated at all nine thresholds.
The systematic variations were propagated at the five thresholds defined
in Section~\ref{sec:reference-ff}; no interpolated ranges are assigned to
the four additional cuts at the quarter and three-quarter positions.
$W/Z$ and other non-jet modelling uncertainties, uncertainties from parton
distribution functions (PDFs), uncertainties from comparisons with alternative
generators, detector uncertainties and the omitted efficiency-scale-factor
uncertainties have not been estimated. The results
therefore provide uncertainty estimates for this Open Data proof of concept rather than a
complete analysis uncertainty budget.
The estimated transfer ranges encompass the inclusive yield discrepancies
at Loose, Medium and Tight and constitute the dominant uncertainty source
estimated in this study.
A full uncertainty estimate for a physics analysis will depend on its
selection and calibration requirements.

\section{Summary and outlook}
\label{sec:discussion}
CUFF estimates jet-to-tau backgrounds by combining four reference fake factors
with candidate-dependent weights. Neural networks learn the reference fake factors at Loose identification and the passing-score distributions from data after correction for non-jet contamination. A separate combination network uses
candidate properties and the target fail-Loose sample to determine the
reference weights. The same trained models provide predictions at
continuously variable Tau-ID thresholds.

The implementation was tested with ATLAS 2015--2016 Open Data. Reference
checks quantify the accuracy and statistical precision of the individual
fake factors. In eight controlled reference mixtures, inclusive yield differences for Medium
identification are $+0.7\%$ to $+4.3\%$, and the extra change introduced by
the learned combination is $-0.62$ to $+0.13$ pp relative
to the data reference.

In the two validation regions, the nominal predictions for Medium identification
underestimate the data references by $21.85\%$ and $45.84\%$.
These differences lie within the broad estimated MC transfer ranges,
whose widths reflect both changes between environments and limited
MC statistics. The nine-threshold results show gradual changes
in the relative differences, demonstrating consistent application of the
same method as the identification requirement is varied.

For Loose identification, CUFF gives an inclusive prediction closer to the data
reference in $N_b=0$ than UFF adapted to public inputs does; the two methods
give more similar inclusive yields in $\mu3j$.
It improves the normalized shapes of several tested distributions in
both regions, including $H_T$, which is not an explicit model input.
The shape results demonstrate a practical advantage of CUFF in this
comparison, alongside the ability to evaluate new
Tau-ID thresholds without retraining.

This study establishes an Open Data proof of concept for CUFF, combining
reference fake factors with learned candidate-dependent weights and
enabling predictions at continuously variable Tau-ID thresholds.
The improved modelling of several distribution shapes and the reuse of
reference measurements motivate applications to differential measurements
and searches involving hadronically decaying tau leptons.

\section*{Acknowledgements}
We thank the ATLAS Collaboration for making the collision and simulated
data used in this study publicly available, together with the associated
software and documentation.

We acknowledge computational support from the High-performance Computing
Public Platform (Shenzhen Campus) of Sun Yat-sen University.

This work was supported in part by the National Key Research and Development
Program of China under Grant No.~2023YFA1605700.

\FloatBarrier

\newcounter{cuffmainfigures}
\newcounter{cuffmaintables}
\setcounter{cuffmainfigures}{\value{figure}}
\setcounter{cuffmaintables}{\value{table}}
\begin{appendices}
\setcounter{figure}{\value{cuffmainfigures}}
\setcounter{table}{\value{cuffmaintables}}
\renewcommand{\thefigure}{\arabic{figure}}
\renewcommand{\thetable}{\arabic{table}}
\numberwithin{equation}{section}
\renewcommand{\theequation}{\Alph{section}\arabic{equation}}

\section{Training and numerical implementation}
\label{app:training}
\label{app:implementation}
\subsection{Score density and training objectives}
\label{app:score-density}
The passing score is rescaled to $u=(s-t_L)/(1-t_L)$. Its density is
\begin{equation}
 \begin{gathered}
 \begin{aligned}
 f_k(u\mid x,I(t_L))={}&\sum_{j=1}^{8}\pi_{kj}(x)\\
 &{}\times\operatorname{Beta}(u;j,9-j),
 \end{aligned}\\
 \pi_{kj}\geq0,\quad\sum_{j=1}^{8}\pi_{kj}=1.
 \end{gathered}
 \label{eq:density}
\end{equation}
The Beta components form a normalized degree-seven Bernstein basis.
The density in the original score is $p_k=f_k/(1-t_L)$.
Integrating $f_k$ above $(t-t_L)/(1-t_L)$ gives the survival fraction
in Eq.~\eqref{eq:intro-cuff-tail}. Equivalently, the stored coefficients
$L_k\pi_{kj}$ can be integrated directly to obtain the FF.
Let $y=1$ for Loose-ID and $y=0$ for $\Azero$. The per-candidate loss is
\begin{equation}
 \begin{aligned}
 \ell_k(x,y,u)={}&\log[1+L_k(x)]\\
 &{}-y\log L_k(x)\\
 &{}-y\log f_k(u\mid x,I(t_L)).
 \end{aligned}
 \label{eq:loss}
\end{equation}
The loss is averaged with the contamination-corrected data weights.
Both categories constrain $L_k$; only passing candidates contribute to
the score-density term. The model output is
$\log L=12\tanh(z_L/12)$, where $z_L$ is the unconstrained scalar
rate output; softmax gives the eight density weights.
For log-density evaluation, $u$ is restricted to $[10^{-7},1-10^{-7}]$; the
nominal training reports indicate that this endpoint correction was not
required.

The combination network adds a four-output correction $h$ to the
reference-classifier log-probabilities.
\begin{equation}
 \begin{gathered}
 g_k(v,C_T)=\\
 \operatorname{softmax}_k\!\left(
 \frac{\log q^{(0)}(v)+h(v,\log q^{(0)}(v),C_T)}{T_{\rm cal}}
 \right),
 \end{gathered}
 \label{eq:head}
\end{equation}
The calibration temperature $T_{\rm cal}$ is selected in $V$.
The training loss is the negative logarithm of the probability assigned
to the candidate's known reference label, with the mixture weighting described in
Section~\ref{sec:reference-combination}. Each training mixture draws
256 candidates per reference. With probability 0.2, the proportions are
chosen from the equal mixture and the four permutations of
$(0.7,0.1,0.1,0.1)$; otherwise a symmetric Dirichlet mixture is drawn,
choosing concentration 0.3, 1 or 3. Each epoch uses 64 training mixtures,
and model selection uses 128 fixed mixtures. The source summaries come
from the opposite sample half and are combined linearly at the chosen
proportions.

\subsection{Data partitions and contamination correction}
\label{app:contamination-implementation}
For evaluation fold $f$, model selection uses $(f+1)\bmod5$ and the other
three folds form $T$. For each inner fold of $T$, the contamination
correction is determined from the other two, with an internal record
split used to select the network. The correction of each $V$ candidate
is the mean of its three estimates. A separate correction model trained
on $T$, selected on $V$, provides correction-only diagnostics on $E$.
Repeated records retain their event-based assignment. The classifier
used by both halves of a reference mixture is trained on the full
permitted $T$ sample.

The full contamination network uses the 17 inputs of
Table~\ref{tab:features}. The score logit is evaluated after restricting
$s$ to $[10^{-6},1-10^{-6}]$. The weighted binary-classification loss
retains the physical data/non-jet-MC yield ratio and signed MC weights.
The output estimates the data-to-non-jet-MC yield-density ratio $r$ as
$\log r=12\operatorname{sigmoid}(a)$, where $a$ is the unconstrained
scalar network output, and gives $c=1-1/r$.
Let $D$ and $M$ denote the permitted training data and non-jet-MC yields.
For nominal training, the full network requires at least 500 data candidates, an
effective non-jet-MC count of at least 50 and $0<M<D$.
It uses two width-16 layers below 3000 training data candidates and two
width-64 layers otherwise. Loose-ID samples with 100--499 candidates
and the same MC requirements use a single width-8 layer with prong and
score-logit inputs. Other cases use
$c=\max[0,\min(1,1-M/D)]$ for each pass/fail category, pooling its one- and three-prong candidates.
For one-prong Loose-ID candidates in MJ-fail, $c=1-M/D$ is determined
from the training yields separately for candidates with and without a
valid associated-jet width, requiring $D>0$ and $0\leq M\leq D$ in each
case.
The $\Azero$ and three-prong weights retain the original treatment.

\subsection{Optimization and small samples}
Table~\ref{tab:expert-training} summarizes the reference FF network configurations.

\begin{table*}[tp]
\centering\small
\caption{Nominal FF network configurations, used in all five outer folds
of each reference region. The hidden-layer column gives the number of
nodes in the two layers. An epoch is one pass through the training sample.
The maximum epoch count limits training; the selected model is the one
with the lowest validation loss.}
\label{tab:expert-training}
\begin{tabular}{lrrr}
\toprule
Reference region & Inputs & Nodes per hidden layer & Maximum epochs\\
\midrule
SS $W$ & 19 & $64,64$ & 5000\\
$t\bar t$ & 17 & $16,16$ & 5000\\
MJ-pass & 16 & $64,64$ & 2000\\
MJ-fail & 14 & $16,16$ & 2000\\
\bottomrule
\end{tabular}
\end{table*}

FF training uses AdamW with learning rate $3\times10^{-4}$, weight decay
$10^{-5}$, batches of 8192 and maximum gradient norm 10. Input
transformations and missing-value replacements are determined from $T$.
The learning rate is halved after 30 epochs without a validation-loss
improvement of $10^{-5}$, with ten epochs between reductions and a
minimum rate of $3\times10^{-6}$. Stopping requires at least 240 epochs,
120 epochs without improvement and two rate reductions. The lowest-loss
parameters are retained.

The effective positive training count, including replica multiplicity
$m_i$ in $w_i$, is
\begin{equation}
 N_{\rm eff}=\frac{(\sum_iw_i)^2}{\sum_{i:m_i>0}w_i^2/m_i}.
 \label{eq:training-neff}
\end{equation}
The full reference Loose-ID training sample selects the capacity for
both prongs: below 100 effective candidates a constant model is used,
from 100 to 999 two width-16 layers are used, and at least 1000 selects
width 64 unless the recorded configuration overrides it. All 20 nominal models used to estimate the reference FFs are neural
networks. Constant rates are stabilized towards the
combined-prong training rate with weight equivalent to 100 $\Azero$
candidates; the score shape uses a prior equivalent to five ID candidates. Their formulas and training configurations are provided in the public
implementation.

The reference classifier and combination network use AdamW with learning
rate $3\times10^{-4}$, weight decay $10^{-4}$ and a 300-epoch limit.
Their validation patience is 30, with minimum durations 30 and 50 epochs,
respectively. Temperature calibration minimizes the weighted validation
cross-entropy over $[0.5,3]$. Classifier probabilities are floored at
$10^{-12}$ only when taking logarithms.

\subsection{MJ-fail rate calibration}
\label{app:mjfail-calibration}
The rate correction uses fixed $p_T$/prong bins with boundaries
20, 25, 30, 40 and 60 GeV and an open final bin. In each bin $b$,
\begin{equation}
 r_b=
 \frac{\sum_{i\in T\cap I(t_L)\cap b}w_i^c}
      {\sum_{i\in T\cap A_0\cap b}w_i^cL_{\rm parent}(x_i)}.
 \label{eq:MJfail-rate}
\end{equation}
Here $w_i^c$ includes the contamination correction and replica
multiplicity, and $L_{\rm parent}$ is the rate before calibration.
The numerator and denominator must be positive, with effective counts of at
least 50 for corrected $\Azero$, 20 for corrected Loose ID and 20 for the
FF-weighted denominator. Otherwise $r_b=1$ and the correction is
marked unmeasured. Multiplying all eight coefficients by $r_b$ preserves
the normalized score shape and continuous threshold dependence.
In each calibrated training bin, applying the calibrated Loose FF to
the corrected $\Azero$ candidates reproduces the corrected Loose yield
by construction. The performance is checked in evaluation samples.
The kinematic bins can introduce steps in the $p_T$ dependence.

\subsection{Target summaries and public inputs}
\label{app:target-summary}
The target moments use signed data-minus-non-jet-MC sums and are not
clipped. The subtraction can place estimated moments outside the
bounds of a positive probability distribution. A nonpositive net
$\Azero$ yield is rejected. The 14 entries describe the first and second
moments of classifier outputs rather than the full target distribution.

\label{app:nnjvt-input}
The stored \texttt{NNJvtPass} decision is used without recalculation.
Its production software version, working point and kinematic range have
not been fully traced for this release. The public tool's default
settings~\cite{NNJvtTool} do not establish the stored configuration. The
associated-jet restriction $20<p_T\leq60$ GeV is retained for this study.

\label{app:threshold-checks}
Predictions at the five thresholds defined in Section~\ref{sec:reference-ff}
were replayed before evaluating the four additional cuts in
Table~\ref{tab:queries}. Nonnegative coefficients and independently
checked Beta tail integrals give the threshold behaviour described in
Section~\ref{sec:application}. Signed data-minus-MC totals can nevertheless
change differently from the individual nonnegative FFs.

\section{Construction of the modelling variations}
\label{app:uncertainty-details}
\subsection{Reference learning responses}
\label{app:learning-response}
The calibration and check samples are disjoint candidate subsets drawn
from reference evaluation samples used during method development.
Candidates in the same reference fold use the same trained models.
Source responses $R/K-1$ and
combination responses $K(h)/P(h)-1$ are measured in groups defined by
prong, NNJvt, tau $p_T$, jet information and exclusive score band.
The source response additionally distinguishes the reference region.
Grouping is fixed in the nominal calibration sample; the complementary
sample is used to check the propagated response.

The support requirements include at least five ID data records, an effective
$\Azero$ count of at least 25, an effective subtracted ID count of at least
four, and positive nominal yields. Prediction yields must exceed
twice their paired-replica uncertainty. For source responses, the
relative uncertainty of the measured ratio $R/Y_{\Azero}$, obtained by
first-order covariance propagation, must not exceed 0.5. Here
$Y_{\Azero}$ is the source $\Azero$ yield after non-jet-MC subtraction,
counted once per distinct kinematic cell when score bands are merged.
For both $R/Y_{\Azero}$ and $R/K$, the standard deviation of the replica
ratios must differ from its linear estimate by no more than 50\% of the
latter. The same stability requirement applies to $K/P$ for the
combination response.
Nonpositive replica denominators lead to merging or a response that
cannot be estimated; replicas are not discarded.

Merging preserves the prong, NNJvt and valid-domain categories;
source-response groups also remain separate by reference region.
The merging order is $p_T$, jet information and score band. The
highest-boundary insufficient group is considered first and merged with
the lower adjacent neighbour when available. Empty cells and missing
categories remain recorded. The grouping is determined from statistical
support, without selecting the largest residual or using the check
sample.

The adopted family contains 19 FF states: the nominal state, an up and
down variation for each of four references and two prongs, and coherent
all-component up and down states. For each state, the combination
response is measured using 18 controlled calibration-mixture choices.
The variation directions for the two prong categories take the nine
combinations in $\{-1,0,+1\}^2$,
giving 3078 joint states. The adopted variations use the measured signed
response $b$ ($b_F$ or $b_g$). A separate diagnostic uses the residual
magnitude $\sqrt{\max(b^2-\widehat{\mathrm{Var}}(b),0)}$.
The FF and conditional combination responses are recalculated in the
same 20 replicas with fixed nominal groups.
When a response cannot be estimated, a factor of one is used to
propagate the other responses and the unmeasured response is flagged.
The separate 256-corner FF sensitivity check is not the adopted joint
family.

\subsection{MC transfer range}
\label{app:migration-algorithm}
The MC comparison uses seven regions: the earlier $Z(\mu\mu)$ selection,
$t\bar t$, MJ-pass, MJ-fail, SS $W$, inclusive OS $W$ and a supplementary
SS high-$m_T$ selection. Each FF is normalized to its response in its own
reference region. The $Z$ and high-$m_T$ samples supply comparison
information only. The primary truth-jet association groups known types
as $q=(u,d,s,c)$, $b$ and $g$; unknown or invalid associations remain
outside the uncertainty estimation. An alternative reconstructed-jet label scheme is a
sensitivity check, not a second independent uncertainty.

The groups distinguish prong, NNJvt status (pass, fail or unknown),
tau-$p_T$ and five mutually exclusive score intervals bounded by the
five thresholds of Section~\ref{sec:reference-ff} and the upper score
limit of one. The final interval contains candidates passing Tight.
The $p_T$
boundaries are 20, 25, 30, 40, 60, 100 and 200 GeV plus overflow.
Leading-jet categories are summed, including absent and missing values.
A fixed record-key hash splits MC into calibration and check samples.
Each reference and comparison group must have at least five ID records, an
effective $\Azero$ count of at least 25, positive ID/prediction/$\Azero$ yields,
and ID and prediction yield-to-error ratios of at least two.
The double ratio must be positive and finite with relative linear
uncertainty at most 0.5, retaining the covariance of signed weighted sums.

Insufficient cells are merged first in neighbouring $p_T$ bins of the
same score band, then in neighbouring score bands. Highest-boundary
insufficient groups are processed first, preferring the lower neighbour.
Complete groups are merged, $\Azero$ is counted once per distinct
$p_T$ cell, and reference, truth, prong and NNJvt categories remain
separate. The groups fixed in the calibration sample are also used for
the check sample. Groups that remain unsupported are recorded as having
no uncertainty estimate.

For each original candidate/score cell $\gamma$, the accepted comparisons
with other regions $e$ and truth types $\zeta$ give lower and upper
score-density factors
\begin{equation}
 \begin{aligned}
 \lambda^-_{k,\gamma}&=\min\{1,D_{k,e,\zeta,\gamma}\}_{\rm supported},\\
 \lambda^+_{k,\gamma}&=\max\{1,D_{k,e,\zeta,\gamma}\}_{\rm supported}.
 \end{aligned}
 \label{eq:migration-field}
\end{equation}
Different comparisons can use different merged groups; their factors
are mapped back to the original cells. They multiply the corresponding
score-density contributions, with the same factor at all leading-jet
values.

For mode $m=(k,p)$, a single down, nominal or up choice is applied to all
candidate and score categories of FF $k$ and prong $p$. The eight modes
have propagated target changes $\Delta P_m^-$ and $\Delta P_m^+$, giving
\begin{equation}
 \begin{aligned}
 \Delta P_{\rm low}
   &=\sum_{m=1}^{8}\min(0,\Delta P_m^-,\Delta P_m^+),\\
 \Delta P_{\rm high}
   &=\sum_{m=1}^{8}\max(0,\Delta P_m^-,\Delta P_m^+).
 \end{aligned}
 \label{eq:migration-envelope}
\end{equation}
The extrema are calculated for each reported bin and threshold.
Different bins may select different mode directions, so the result is
not a simultaneous confidence band. The signed data and non-jet-MC
contributions and their existing reference weights are retained.
Figure~\ref{fig:migration} shows the ranges from the MC transfer component,
while Table~\ref{tab:uncertainty-components} reports the low-NNJvt data
variation separately. In the combined MC/low-NNJvt prescription, the
low-NNJvt data response replaces the MC response for the MJ-fail FF
component of candidates failing NNJvt.

Categories without an uncertainty estimate use a numerical factor of
one. For Medium identification under the combined prescription, these categories account
for about 0.13\% and 3.75\% of the FF-weighted $\Azero$ data yield
before MC subtraction in $N_b=0$ and $\mu3j$, respectively. These are
weighted-yield fractions, not bounds on other missing modelling effects.

\subsection{Non-jet MC variations}
\label{app:nonjet-variations}
The non-jet estimate enters reference-training corrections, the target
summary, the FF-weighted $\Azero$ subtraction and the ID reference.
The same MC variation is used in each affected calculation, including
model training. Luminosity variations rescale all non-jet-MC
contributions; the $t\bar t$ cross-section variation changes only that
process. The public luminosity uncertainties and
$\sigma_{t\bar t}=832^{+46}_{-52}$ pb are used~\cite{Luminosity,TTNormalization}.
The combined luminosity variation is approximately 0.87\%.

The representative $t\bar t$ modelling study uses six renormalization-
and factorization-scale weight choices and two Var3c radiation weights
from the public samples 410470 and 410471. Each alternative weight set
is normalized by its corresponding full-sample generator-weight sum
with the inclusive cross-section parameter fixed. The study therefore
changes shapes and selection acceptance without repeating the
cross-section normalization variation. The nominal and varied models
use the same training settings. Their prediction difference gives the
shift in $P$. The matching ID non-jet yield is also varied when
calculating the change in $P-R$. The scale and radiation components are
reported separately. The nominal statistical and training repetitions
are retained separately from these generator-weight variations.

For Medium identification, the shifts in $P-R$, expressed in percentage points relative
to the nominal ID reference $R_0$, are $[-0.52,+0.00]$ in $N_b=0$ and
$[-7.93,+8.36]$ in $\mu3j$ for the normalization variations. The corresponding ranges
are $[-0.035,+0.027]$ and $[-0.113,+1.526]$ for the $t\bar t$ scale
variations, and $[-0.055,+0.005]$ and $[+0.000,+1.705]$ for the
radiation variations, respectively.
These non-jet systematic variations are not included in the statistical
and training error bars in Section~\ref{sec:results}, or in the
MC-transfer-only ranges. The small effects of the tested $t\bar t$
variations do not determine the size of the non-jet modelling
uncertainties that have not been estimated.

\section{Additional validation details}
\label{app:supplementary-checks}
\subsection{Resampling and coverage checks}
\label{app:statistical-details}
\label{app:toy}
The common resampling keys are (year, run, event) for data and
(DSID, run, event) for MC. Original file, entry and candidate identifiers
are checked separately because some MC samples reuse run/event keys.
Records with the same key share a resampling multiplicity. The $t\bar t$
and MJ-pass replicas include execution and optimization differences, so
their spread is interpreted as statistical and training variation rather
than a separate hardware uncertainty.

The sample covariance is calculated from the twenty repetitions with
denominator 19 and therefore has rank at most 19. Each repetition is
first summed over the five evaluation folds. These covariance details
apply to the statistical and training uncertainties in
Section~\ref{sec:uncertainties}.

The coverage study uses 100 independent pseudo-experiments. Each trains
four FF networks, a reference classifier and a combination network,
followed by 20 joint Poisson-resampling repetitions. Five preselected
experiments have 80 further repetitions, giving 2500 sets of six models
and 15\,000 training runs in total. No failed experiment or nonpositive
reference was omitted. Sixteen input variables are functions of two
underlying coordinates, a scalar and the prong category. The generating
score density uses the same eight-component Beta basis as the FF
networks, with known reference purities and no physical mismatch between
regions.

The fixed source proportions are
$\boldsymbol\alpha_A=(0.4,0.2,0.3,0.1)$ and
$\boldsymbol\alpha_B=(0.1,0.55,0.15,0.2)$, at expected fake-$\Azero$
counts 600 and 6000. Five cumulative thresholds and four adjacent score
intervals are checked for both prongs and their sum. Nominal 95\%
intervals use half-width $1.959964s$, where $s$ is the paired-replica SD.
The two primary checks use different comparison quantities.
The finite-target check constructs both $P$ and $R$ from sampled target
events and tests whether the interval for $P-R$ contains zero. For
mixture $A$ with 600 expected fake-$\Azero$ events, the three-prong
$MT_{1/2}$-to-Tight score interval has 84\% coverage.

The separate learning check removes finite-target fluctuations.
For each trained model set, the target summary $C_T$ is calculated by
averaging the reference-classifier outputs and their products over the
known generating distribution. The combined FF is then averaged over
the same distribution, using the 32 feature/prong cells and their known
probabilities, and multiplied by the expected fake-$\Azero$ yield.
The resulting prediction is compared with the known expected ID yield.
Its uncertainty is the standard deviation of these averaged predictions
across the twenty resampled and retrained reference-model sets.
Thus the target distribution is known exactly in this check, while the
reference training samples remain finite.

For mixture $A$ with 600 expected fake-$\Azero$ events, the three-prong
Medium-to-$MT_{1/2}$ interval in this second check has bias 0.08832,
empirical SD 0.07162 and RMS resampling SD 0.07374, in expected-event
units, with 78\% coverage. The estimated spread is close to the observed
spread, but the central prediction is biased. This illustrates why an
uncertainty describing statistical fluctuations need not cover a
learning bias.

\subsection{Controlled reference mixtures}
\label{app:controlled-evidence}
For the eight evaluation mixtures, candidates from reference $k$ receive
the coefficient $\beta_k/(5Y^{E,0}_{k,\Azero})$, where
$Y^{E,0}_{k,\Azero}$ is the nominal signed $\Azero$ yield in its complete
evaluation fold. The same coefficient is applied to ID and $\Azero$
candidates and held fixed across replicas. The nominal full mixture
then has the stated anti-ID proportions; each half, prong category or
replica need not. For prediction, the two halves exchange candidate and
summary roles. The separate learning-response calibration mixtures use
reference-training normalizations, as specified in their configuration.

\subsection{Adapted-UFF comparison}
\label{app:comparison-details}
For candidate $i$ in tau-$p_T$/prong bin $b(i)$, the adapted FF is
\begin{equation}
 F_i^{\rm UFF\text{-}style}=\sum_{k=1}^{4}
       \alpha_{k,b(i)}F^{\rm bin}_{k,b(i)}.
 \label{eq:UFFproxy}
\end{equation}
Here $F^{\rm bin}_{k,b}$ is the binned reference FF and $\alpha_{k,b}$
is its template coefficient. The fit uses signed, normalized
data-minus-non-jet-MC histograms and
iterated weighted least squares, including template uncertainties and
target covariance. It requires effective counts of at least 50 in each reference template and at
least 100 in the target, and a condition number below $10^8$. Otherwise it combines $p_T$ bins within a prong category and
then both prongs. An unresolved case uses equal coefficients and retains
its flag. Separate fixed rules merge insufficient reference-FF bins.
The neural MJ-fail rate calibration is not applied to the directly measured binned fake factors.

The comparison uses the same event-fold assignments as CUFF. For each
target fold, the template coefficients are fitted using the target
$\Azero$ events in the other four folds and templates from the
corresponding reference training sample. The coefficients are then
applied to candidates in that target fold.

All 50 nominal fold/bin fits in each region satisfy the
numerical requirements. Of these, 40 use individual bins and ten use
merged $p_T$ bins. At least one fitted coefficient is negative in each
fit. Negative candidate FFs occur in 4598 $N_b=0$ records,
with an absolute weighted contribution of about 3.81 events, and none
in nominal $\mu3j$. The conditional
WLS $\chi^2/n_{\rm dof}$ medians are 4.65 and 3.49, with maxima 23.30
and 16.42. These values describe remaining template differences and are
not calibrated $p$-values. The fits, templates and FFs are recalculated
in each common replica; their counting and model fluctuations are not
added again as independent terms.

\subsection{Supplementary numerical results}
\label{app:shape-details}
The public reproducibility package contains all 36
reference-region Loose/Medium/Tight results, all 72 controlled-mixture
results and their component covariances and replica-mean shifts, and
all 48 marginal/coarse-joint shape comparisons. The source tables are:
\begin{quote}
\raggedright\small
\texttt{source\_LMT36.csv}\\
\texttt{controlled\_mixtures\_all72.csv}\\
\texttt{multivariable\_shapes\_Loose.csv}
\end{quote}
The estimated uncertainty components at the standard thresholds are in:
\begin{quote}
\raggedright\small
\texttt{uncertainty\_components\_LMT.csv}
\end{quote}

The shape comparisons use fixed bins, retain signed contents and all
flow categories, and normalize each replica separately. Their method
uncertainties describe paired differences rather than independent-error
addition or calibrated significances. The coarse joint comparison uses
tau $p_T$, prong and jet-count categories (0, 1 or at least 2); in
$\mu3j$ only the last jet-count category is populated. The data-and-code availability statement provides
links to the public reproducibility package.

\FloatBarrier

\begingroup
\makeatletter
\renewenvironment{table*}[1][]{%
  \par\begingroup\def\@captype{table}\centering\small
}{\par\endgroup\vspace{16pt}}
\makeatother
\twocolumn[{%
\begin{minipage}{\textwidth}
\section{Numerical validation results}
\label{app:validation-tables}
Tables~\ref{tab:OS} and~\ref{tab:added} give the validation yields and
relative differences discussed in Sections~\ref{sec:OS}
and~\ref{sec:nine-thresholds}. Tables~\ref{tab:UFFcomparison}
and~\ref{tab:multivariable-shapes} give the adapted-UFF and shape comparisons
of Sections~\ref{sec:UFFresult} and~\ref{sec:shape-results}.
The definitions and uncertainty conventions are specified in the captions.
\par\medskip

\begin{table*}[tp]
\centering\small
\caption{Predicted and reference yields in the two validation regions at the standard Tau-ID working points, combining one-prong and three-prong candidates. $P_0$ is the nominal CUFF prediction and $R_0$ the data yield in the ID region after subtraction of simulated non-jet contributions. The final columns give $d=100(P_0-R_0)/R_0$ and its statistical and training uncertainty $s_d$. Systematic-variation ranges are reported separately.}
\label{tab:OS}
\begin{tabular}{llrrrr}
\toprule
Validation region & Tau-ID & Prediction $P_0$ & Reference $R_0$ & $d$ [\%] & $s_d$ [pp] \\
\midrule
$N_b=0$ & Loose & 269,066.9 & 335,132.2 & -19.71 & 1.06 \\
$N_b=0$ & Medium & 144,726.1 & 185,180.2 & -21.85 & 1.58 \\
$N_b=0$ & Tight & 66,545.2 & 90,450.8 & -26.43 & 2.01 \\
$\mu3j$ & Loose & 3,078.7 & 6,181.1 & -50.19 & 4.49 \\
$\mu3j$ & Medium & 1,718.2 & 3,172.4 & -45.84 & 6.03 \\
$\mu3j$ & Tight & 846.5 & 1,424.0 & -40.56 & 8.67 \\
\bottomrule
\end{tabular}
\end{table*}

\begin{table*}[tp]
\centering\small
\caption{Inclusive yield differences at nine Tau-ID thresholds, combining
prong categories. The relative difference $d$ from ID data after non-jet
subtraction is given in percent, and its statistical and training
uncertainty $s_d$ in pp. Threshold results are correlated through shared
events and models. Systematic uncertainties are not included.}
\label{tab:added}
\setlength{\tabcolsep}{3pt}
\begin{tabular}{lrrrr}
\toprule
 & \multicolumn{2}{c}{$N_b=0$} & \multicolumn{2}{c}{$\mu3j$} \\
\cmidrule(lr){2-3}\cmidrule(lr){4-5}
Threshold & $d$ [\%] & $s_d$ [pp] & $d$ [\%] & $s_d$ [pp] \\
\midrule
Loose & -19.71 & 1.06 & -50.19 & 4.49 \\
$LM_{1/4}$ & -19.96 & 1.18 & -48.60 & 4.87 \\
$LM_{1/2}$ & -20.40 & 1.32 & -47.36 & 5.34 \\
$LM_{3/4}$ & -21.14 & 1.46 & -46.18 & 5.79 \\
Medium & -21.85 & 1.58 & -45.84 & 6.03 \\
$MT_{1/4}$ & -23.08 & 1.76 & -45.32 & 6.43 \\
$MT_{1/2}$ & -24.39 & 1.87 & -44.70 & 7.10 \\
$MT_{3/4}$ & -25.51 & 1.97 & -42.13 & 7.70 \\
Tight & -26.43 & 2.01 & -40.56 & 8.67 \\
\bottomrule
\end{tabular}
\end{table*}

\begin{table*}[tp]
\centering\small
\caption{Loose-ID comparison of CUFF and adapted UFF in the two validation regions. Each method column gives the relative difference $d$ in percent and its statistical uncertainty in percentage points, including template-fit and training fluctuations. The final column gives $100(P_{\rm UFF}-P_{\rm CUFF})/R_0$ and its paired standard deviation, both in pp. The reference is data in the ID region after non-jet subtraction. The adapted UFF inputs are defined in Section~\ref{sec:UFFmethod}; the comparison does not include complete systematic uncertainty budgets.}
\label{tab:UFFcomparison}
\begin{tabular}{llrrr}
\toprule
Validation region & Prongs & CUFF & Adapted UFF & Adapted UFF $-$ CUFF \\
\midrule
$N_b=0$ & 1+3 & $-19.71\pm1.06$ & $-55.14\pm1.51$ & $-35.42\pm1.75$ \\
$N_b=0$ & 1 & $-20.52\pm1.39$ & $-57.29\pm1.73$ & $-36.77\pm2.09$ \\
$N_b=0$ & 3 & $-17.34\pm1.17$ & $-48.80\pm2.26$ & $-31.46\pm2.80$ \\
$\mu3j$ & 1+3 & $-50.19\pm4.49$ & $-43.52\pm2.21$ & $+6.67\pm5.28$ \\
$\mu3j$ & 1 & $-51.01\pm5.38$ & $-50.17\pm2.60$ & $+0.83\pm5.81$ \\
$\mu3j$ & 3 & $-46.99\pm7.23$ & $-17.49\pm4.58$ & $+29.51\pm7.12$ \\
\bottomrule
\end{tabular}
\end{table*}

\end{minipage}
}]
\twocolumn[{%
\begin{minipage}{\textwidth}
\noindent\textbf{Appendix D: Numerical validation results (continued)}
\par\medskip

\begin{table*}[tp]
\centering\small
\caption{Loose-ID shape comparison after normalizing each prediction and the data reference by its own total, combining prong categories. Smaller $D_{\rm shape}$, defined in Eq.~\eqref{eq:shape-distance}, indicates closer shapes. All entries are in percentage points. The method difference $\Delta D=D^{\rm UFF}_{\rm shape}-D^{\rm CUFF}_{\rm shape}$ is followed by its standard deviation from the 20 paired statistical and training replicas. The four observables below the divider are not explicit model inputs. Absolute-yield predictions remain unchanged; a full systematic comparison is not included.}
\label{tab:multivariable-shapes}
\begin{tabular}{lrrrrrr}
\toprule
& \multicolumn{3}{c}{$N_b=0$} & \multicolumn{3}{c}{$\mu3j$} \\
Observable & CUFF & Adapted UFF & $\Delta D$ & CUFF & Adapted UFF & $\Delta D$ \\
\midrule
$p_T^\tau$ & 0.96 & 8.12 & $+7.16\pm 1.21$ & 4.01 & 11.23 & $+7.22\pm 2.94$ \\
$E_T^{\rm miss}$ & 0.66 & 4.88 & $+4.22\pm 0.60$ & 3.23 & 3.64 & $+0.42\pm 2.11$ \\
$N_{\rm jets}$ & 1.64 & 5.43 & $+3.79\pm 0.47$ & 6.46 & 5.98 & $-0.48\pm 2.01$ \\
\midrule
$m_T(\mu,E_T^{\rm miss})$ & 0.37 & 1.25 & $+0.88\pm 0.48$ & 2.41 & 3.63 & $+1.22\pm 0.96$ \\
$H_T$ & 1.05 & 5.35 & $+4.30\pm 0.76$ & 7.53 & 13.10 & $+5.57\pm 1.49$ \\
$\Delta R(\mu,\tau)$ & 1.50 & 1.27 & $-0.23\pm 0.39$ & 1.97 & 2.11 & $+0.14\pm 1.71$ \\
$\eta^\mu$ & 0.90 & 0.78 & $-0.12\pm 0.60$ & 4.91 & 4.90 & $-0.02\pm 1.25$ \\
\bottomrule
\end{tabular}
\end{table*}

\end{minipage}
}]
\endgroup

\end{appendices}

\backmatter
\section*{Data and code availability}
The collision data are the public ATLAS 2015 and 2016 research releases on
the CERN Open Data Portal~\cite{OD2015,OD2016}. The 105-sample nominal MC
selection contains 74 electroweak samples~\cite{MCEW}, 13 QCD samples~\cite{MCQCD},
17 samples from the nominal top record~\cite{MCTop}, and DSID 700000 from the
top-variation record~\cite{MCTopVariation}. DSID 700000 is Sherpa $t\bar tW$;
the older $t\bar tW$ alternative 410155 is excluded. It is not a second
inclusive $t\bar t$ prediction alongside the complementary samples 410470
and 410471.

The analysis code, trained model parameters and numerical results are
publicly available from the \href{https://gitlab.cern.ch/liuya/cuff-reproducibility/-/tree/v0.1.0}{GitLab repository}
and archived on Zenodo~\cite{CUFFReproducibility}. The package includes
public-data inventories, standalone inference examples and scripts to
reproduce the figures.

\section*{Statements and Declarations}
\noindent\textbf{Funding.}
This work was supported in part by the National Key Research and Development
Program of China under Grant No.~2023YFA1605700.

\noindent\textbf{Competing interests.}
The authors declare that they have no competing interests relevant to this work.

\noindent\textbf{Author contributions.}
Yang Liu developed the method. Lei Guo and Yang Liu prepared the manuscript
draft. Antonio De Maria reviewed and revised the manuscript.
Chengxin Liao, Zihang Liu, Wenyi Jin, Jiarong Yuan and Yuchen Cai contributed
to the analysis and validation. Da Xu, Xuai Zhuang, Yanjun Tu and Yang Liu
supervised the study.

\section*{Use of computational assistance}
AI-based tools assisted with implementation, verification and manuscript
preparation. The authors are responsible for the analysis, interpretation
and final text.

\atColsBreak{\vskip\baselineskip}


\begin{thebibliography}{99}
\raggedright
\interlinepenalty=10000 
\bibitem{PDGTau} S.~Navas et al. (Particle Data Group),
Review of Particle Physics, Phys. Rev. D \textbf{110} (2024) 030001,
\url{https://doi.org/10.1103/PhysRevD.110.030001}; tau-lepton listing,
\url{https://pdg.lbl.gov/2024/listings/rpp2024-list-tau.pdf}.
\bibitem{HiggsTau} ATLAS Collaboration,
Measurements of Higgs boson production cross-sections in the
$H\to\tau^+\tau^-$ decay channel in $pp$ collisions at $\sqrt{s}=13$ TeV
with the ATLAS detector, JHEP \textbf{08} (2022) 175,
\url{https://doi.org/10.1007/JHEP08(2022)175},
\href{https://arxiv.org/abs/2201.08269v2}{arXiv:2201.08269v2}.
\bibitem{ChargedHiggsTau} ATLAS Collaboration,
Search for charged Higgs bosons decaying via $H^{\pm}\to\tau^{\pm}\nu_{\tau}$
in the $\tau$+jets and $\tau$+lepton final states with 36 fb$^{-1}$ of
$pp$ collision data recorded at $\sqrt{s}=13$ TeV with the ATLAS experiment,
JHEP \textbf{09} (2018) 139,
\url{https://doi.org/10.1007/JHEP09(2018)139},
\href{https://arxiv.org/abs/1807.07915v2}{arXiv:1807.07915v2}.
\bibitem{UFF} ATLAS Collaboration,
Estimation of backgrounds from jets misidentified as $\tau$-leptons using the
Universal Fake Factor method with the ATLAS detector,
Eur. Phys. J. C \textbf{85} (2025) 1441,
\url{https://doi.org/10.1140/epjc/s10052-025-14916-1},
\href{https://arxiv.org/abs/2502.04156v2}{arXiv:2502.04156v2}.
\bibitem{FakeLeptonTools} ATLAS Collaboration,
Tools for estimating fake/non-prompt lepton backgrounds with the ATLAS
detector at the LHC, JINST \textbf{18} (2023) T11004,
\url{https://doi.org/10.1088/1748-0221/18/11/T11004},
\href{https://arxiv.org/abs/2211.16178v2}{arXiv:2211.16178v2}.
\bibitem{NFF} J.~Gavranovi\v{c} et al.,
Neural Fake Factor Estimation Using Data-Based Inference,
\href{https://arxiv.org/abs/2511.06972v2}{arXiv:2511.06972v2}.
\bibitem{JVT} ATLAS Collaboration,
Performance of pile-up mitigation techniques for jets in $pp$ collisions at
$\sqrt{s}=8$ TeV using the ATLAS detector,
Eur. Phys. J. C \textbf{76} (2016) 581,
\url{https://doi.org/10.1140/epjc/s10052-016-4395-z},
\href{https://arxiv.org/abs/1510.03823v2}{arXiv:1510.03823v2}.
\bibitem{QGTagger} ATLAS Collaboration,
Performance and calibration of quark/gluon-jet taggers using 140 fb$^{-1}$
of $pp$ collisions at $\sqrt{s}=13$ TeV with the ATLAS detector,
Chin. Phys. C \textbf{48} (2024) 023001,
\url{https://doi.org/10.1088/1674-1137/acf701},
\href{https://arxiv.org/abs/2308.00716v2}{arXiv:2308.00716v2}.
\bibitem{MoE} R.~A.~Jacobs, M.~I.~Jordan, S.~J.~Nowlan and G.~E.~Hinton,
Adaptive Mixtures of Local Experts,
Neural Computation \textbf{3} (1991) 79--87,
\url{https://doi.org/10.1162/neco.1991.3.1.79},
\href{https://www.cs.toronto.edu/~hinton/absps/jjnh91.pdf}{public author copy}.
\bibitem{DeepSets} M.~Zaheer et al., Deep Sets,
Advances in Neural Information Processing Systems \textbf{30} (2017),
\href{https://arxiv.org/abs/1703.06114v3}{arXiv:1703.06114v3}.
\bibitem{ClassifierRatio} K.~Cranmer, J.~Pavez and G.~Louppe,
Approximating Likelihood Ratios with Calibrated Discriminative Classifiers,
\href{https://arxiv.org/abs/1506.02169v2}{arXiv:1506.02169v2}.
\bibitem{MUFFIN} CMS Collaboration,
A machine-learning method for the estimation of the background from jets
misidentified as hadronic tau lepton decays,
\href{https://arxiv.org/abs/2610.13634v1}{arXiv:2610.13634v1}.
\bibitem{OD2015} ATLAS Collaboration, 2015 collision data for research,
CERN Open Data record 80000,
\url{https://doi.org/10.7483/OPENDATA.ATLAS.AOQL.8TT3}.
\bibitem{OD2016} ATLAS Collaboration, 2016 collision data for research,
CERN Open Data record 80001,
\url{https://doi.org/10.7483/OPENDATA.ATLAS.4ZES.DJHA}.
\bibitem{MuonRecoRun2} ATLAS Collaboration,
Muon reconstruction and identification efficiency in ATLAS using the full
Run 2 $pp$ collision data set at $\sqrt{s}=13$ TeV,
Eur. Phys. J. C \textbf{81} (2021) 578,
\url{https://doi.org/10.1140/epjc/s10052-021-09233-2},
\href{https://arxiv.org/abs/2012.00578v2}{arXiv:2012.00578v2}.
\bibitem{ElectronRecoRun2} ATLAS Collaboration,
Electron and photon efficiencies in LHC Run 2 with the ATLAS experiment,
JHEP \textbf{05} (2024) 162,
\url{https://doi.org/10.1007/JHEP05(2024)162},
\href{https://arxiv.org/abs/2308.13362v2}{arXiv:2308.13362v2}.
\bibitem{TauDeepSetConfig} ATLAS Offline Software,
Tau-identification configuration in Athena release 22.2.110,
\texttt{TauToolHolder.py},
\url{https://gitlab.cern.ch/atlas/athena/-/blob/release/22.2.110/Reconstruction/tauRec/python/TauToolHolder.py},
accessed 9 November 2026.
\bibitem{TauRun3Performance} ATLAS Collaboration,
Reconstruction, Identification, and Calibration of hadronically decaying
tau leptons with the ATLAS detector for the LHC Run 3 and reprocessed Run 2 data,
ATL-PHYS-PUB-2022-044 (2022),
\url{https://cds.cern.ch/record/2827111}.
\bibitem{NNJvtTool} ATLAS Offline Software,
CP::NNJvtSelectionTool and CP::JvtSelectionToolBase,
public source documentation,
\url{https://atlas-sw-doxygen.web.cern.ch/atlas-sw-doxygen/atlas_main--Doxygen/docs/html/d4/d59/classCP_1_1NNJvtSelectionTool.html},
accessed 26 October 2026.
\bibitem{ParticleFlowJets} ATLAS Collaboration,
Jet reconstruction and performance using particle flow with the ATLAS Detector,
Eur. Phys. J. C \textbf{77} (2017) 466,
\url{https://doi.org/10.1140/epjc/s10052-017-5031-2},
\href{https://arxiv.org/abs/1703.10485v2}{arXiv:1703.10485v2}.
\bibitem{DL1dTagging} ATLAS Collaboration,
Transforming jet flavour tagging at ATLAS,
Nature Commun. \textbf{17} (2026) 541,
\url{https://doi.org/10.1038/s41467-025-65059-6},
\href{https://arxiv.org/abs/2505.19689v2}{arXiv:2505.19689v2}.
\bibitem{AdamW} I.~Loshchilov and F.~Hutter,
Decoupled Weight Decay Regularization, ICLR (2019),
\href{https://arxiv.org/abs/1711.05101v3}{arXiv:1711.05101v3}.
\bibitem{Luminosity} ATLAS Collaboration,
Luminosity determination in $pp$ collisions at $\sqrt{s}=13$ TeV using the
ATLAS detector at the LHC, Eur. Phys. J. C \textbf{83} (2023) 982,
\url{https://doi.org/10.1140/epjc/s10052-023-11747-w},
\href{https://arxiv.org/abs/2212.09379v2}{arXiv:2212.09379v2}.
\bibitem{TTNormalization} ATLAS Collaboration,
Search for heavy particles decaying into top-quark pairs using lepton-plus-jets
events in proton--proton collisions at $\sqrt{s}=13$ TeV with the ATLAS detector,
Eur. Phys. J. C \textbf{78} (2018) 565,
\url{https://doi.org/10.1140/epjc/s10052-018-5995-6},
\href{https://arxiv.org/abs/1804.10823v2}{arXiv:1804.10823v2}.
\bibitem{MCEW} ATLAS Collaboration,
ATLAS DAOD\_PHYSLITE format MC simulation electroweak boson nominal samples,
CERN Open Data Portal, record 80010 (2024),
\url{https://doi.org/10.7483/OPENDATA.ATLAS.K5SU.X65Y}.
\bibitem{MCQCD} ATLAS Collaboration,
ATLAS DAOD\_PHYSLITE format MC simulation QCD jet nominal samples,
CERN Open Data Portal, record 80014 (2024),
\url{https://doi.org/10.7483/OPENDATA.ATLAS.OXQR.DJQ3}.
\bibitem{MCTop} ATLAS Collaboration,
ATLAS DAOD\_PHYSLITE format MC simulation top nominal samples,
CERN Open Data Portal, record 80017 (2024),
\url{https://doi.org/10.7483/OPENDATA.ATLAS.MM1Y.O0PH}.
\bibitem{MCTopVariation} ATLAS Collaboration,
ATLAS DAOD\_PHYSLITE format MC simulation top systematic variation samples,
CERN Open Data Portal, record 80018 (2024),
\url{https://doi.org/10.7483/OPENDATA.ATLAS.F7P0.VAST}.
\bibitem{CUFFReproducibility} L.~Guo et al.,
CUFF reproducibility materials: data catalogue, analysis code and trained models,
version 0.1.0, Zenodo (2026),
\url{https://doi.org/10.5281/zenodo.24455399}.
\end{thebibliography}
\end{document}